\documentclass[12pt,a4paper]{article}
\pdfoutput=1
\usepackage{jheppub}
 \usepackage{amsmath}
 \usepackage{amsfonts}
 \usepackage{mathtools}
 \usepackage{amssymb}
 \usepackage{caption}
 \usepackage{subcaption}
 \usepackage{slashed}
 \usepackage[utf8]{inputenc}
 \makeatletter
\newcommand*{\rom}[1]{\expandafter\@slowromancap\romannumeral #1@}
\makeatother
\title{Shear sum rule in higher derivative gravity theories}
 \author{Subham Dutta Chowdhury}
\affiliation{Centre for High Energy Physics, Indian Institute of Science,\\
C. V. Raman Avenue, Bangalore 560012, India.\\
}
\emailAdd{subhamdutta@iisc.ac.in}
\abstract{ We study holographic shear sum rules in Einstein gravity with curvature squared corrections. Sum rules relate weighted integral over spectral densities of retarded correlators in the shear channel to the one point functions of the CFTs. The proportionality constant can be written in terms of the data of three point functions of the stress tenors of the CFT ($t_2$ and $t_4$). For CFTs dual to two derivative Einstein gravity, this proportionality constant is just $\frac{d}{2(d+1)}$. This has been verified by a direct holographic computation of the retarded correlator for Einstein gravity in $AdS_{d+1}$ black hole background. We compute corrections to the holographic shear sum rule in presence of higher derivative corrections to the Einstein-Hilbert action. We find agreement between the sum rule obtained from a general CFT analysis and holographic computation for Gauss Bonnet theories in $AdS_5$ black hole background. We then generalize the sum rule for arbitrary curvature squared corrections to Einstein-Hilbert action in $d\geq 4$. Evaluating the parameters $t_2$ and $t_4$ for the possible dual CFT in presence of such curvature corrections, we find an agreement with the general field theory derivation to leading order in coupling constants of the higher derivative terms.}

\begin{document}
  \maketitle
\section{Introduction}
          \paragraph{} Sum rules are analytic structures that are used to constrain spectral densities of any quantum field theory \citep{Romatschke:2009ng, Gulotta:2010cu, David:2011hy, Chowdhury:2016hjy, Witczak-Krempa:2015pia, WitczakKrempa:2012gn, Myers:2016wsu, Katz:2014rla}. Schematically they are weighted moments of the spectral density over frequency, which are proportional to one point functions in the theory. Such non perturbative constraints on the spectral density first arose in the context of QCD, where it is easier to obtain one point functions rather than the full thermal correlator. In the context of $\mathcal{N}=4$ SYM at finite temperature and zero chemical potential, this was first studied in \cite{Romatschke:2009ng}. The shear sum rule for spectral density corresponding to the retarded correlator $T_{xy}$ was formulated. 
 \begin{eqnarray}\label{n4sym}
 \frac{2}{5} \epsilon = \frac{1}{\pi} \int_{-\infty}^{\infty}
  \frac{d\omega}{\omega} \left[ \rho (\omega) - \rho_{T=0}(\omega) \right]
  ,
\end{eqnarray} 
where 
\begin{eqnarray}
\rho(\omega)= \textrm{Im} G_R(\omega), 
\end{eqnarray} 
is the spectral density corresponding to the retarded shear correlator defined  in 
position space by 
\begin{equation}
G_R(t, \vec x)=  i\theta(t)\langle [T_{xy}(t, \vec x),T_{xy}(0)] \rangle.
\end{equation}          
The expectation values have been taken over states held at finite temperature and zero chemical potential. Using conformal invariance and low energy hydrodynamic behaviour , the LHS of the sum rule is determined in terms of the parameters of the three point functions of the stress tensor. Moreover, for CFTs with two derivative gravity duals, $AdS/CFT$ provides an useful prescription for computing retarded correlators of conserved currents at strong coupling. Authors of \cite{Romatschke:2009ng} have also verified the shear sum rule \eqref{n4sym} at strong coupling by directly computing the retarded correlator $G_R(\omega)$ in an $AdS_5$ black hole background. From the holographic side, one studies fluctuations which are dual to the stress tensor in the boundary CFT. The fluctuations obey the equations of motion of a minimally coupled scalar. The shear sum rule is controlled by the high frequency behaviour of the retarded correlator computed in this manner. The holographic calculation is an independent check of the sum rule.   
\paragraph{ }  Shear sum rules for arbitrary $d$ dimensional CFTs has been recently computed in \cite{Chowdhury:2016hjy}. 
\begin{eqnarray}\label{introdsumrule}
 \lim_{\epsilon \rightarrow 0^+}\frac{1}{\pi}\int_{-\infty}^{\infty} \frac{\delta\rho(\omega)d\omega}{\omega -i\epsilon} &=&  \left(\frac{(-1+d) d}{2 (1+d)}+\frac{(3-d) t_2}{2 (-1+d)}+\frac{\left(2+3 d-d^2\right) t_4}{(-1+d) (1+d)^2}\right)\frac{\epsilon}{d-1}, \nonumber\\
\rho(z) &=& \textrm{Im} G_R(\omega),\nonumber\\
G_R(t,x)&=& i\theta(t)[T_{xy}(t,x),T_{xy}(0)],\nonumber\\
\delta\rho(\omega)&=& \rho(\omega) -\rho(\omega)_{T=0}, \nonumber\\
\end{eqnarray}
where $\epsilon$ refers to the energy density of the theory and 
$t_2,t_4$  are the linearly independent parameters that appear in the conformal collider formalism of Hofman and Maldacena  \cite{Hofman:2008ar}. These dimensionless constants $t_2$ and $t_4$ are functions of the parameters $a, b, c$  of three point function of the stress tensor of the CFTs \cite{Osborn:1993cr}. Note that this sum rule , which is derived purely based on conformal invariance and low energy hydrodynamic behaviour, is valid for any coupling for any CFT. For CFTs with two derivative gravity dual, $t_2, t_4=0$. 
\begin{eqnarray}\label{egsumrule}
  \frac{ d}{2 (1+d)}\epsilon &=& \lim_{\epsilon \rightarrow 0^+}\frac{1}{\pi}\int_{-\infty}^{\infty} \frac{\delta\rho(\omega)d\omega}{\omega -i\epsilon}. \nonumber\\
\end{eqnarray}
This agrees with the conclusions of \citep{Gulotta:2010cu}, where the shear sum rule was computed holographically for theories with $AdS_{d+1}$ duals. The analyticity of the retarded correlator, which is an assumption for obtaining the sum rule, was argued for by analysing the differential equation corresponding to the fluctuations in the shear channel. Modifications to the holographic shear sum rule in presence of a chemical potential has also been studied \citep{David:2012cd}. In all the previous analysis, only the leading term in \eqref{introdsumrule} has been verified. We want to check the coefficient of $t_2$ from holography.

\paragraph{ } In this paper we study the modification to the shear sum rule \eqref{egsumrule}, due to the presence of curvature squared corrections to Einstein-Hilbert action. Higher derivative corrections to two derivative gravity arise because of stringy or quantum corrections of the classical action \cite{Gubser:1998nz}. In terms of possible dual CFT, this correspond to possible $\frac{1}{\lambda}$ or $\frac{1}{N_c}$ corrections. For example, the leading order corrections in $\frac{1}{\lambda}$ to $\mathcal{N}=4$ SYM correspond to stringy corrrections of the type \rom{2}B supergravity on $AdS_5 \times S^5$, usually of the form $\alpha'^3 R^4$. There has been a systematic study of effects of such higher derivative corrections to the ratio of shear viscosity $\eta$ to the entropy density $s$ \cite{Buchel:2008ae, Buchel:2008vz, Buchel:2009sk, Buchel:2009tt}. We look at corrections to the Einstein-Hilbert action in an $AdS$ background that terminate at fourth order in derivative expansion and assume that there exists some $d$ dimensional CFT at the boundary. 
\begin{eqnarray}\label{introhdaction}
S=\int d^D x \sqrt{-g} \left( \frac{R}{2\kappa}- \Lambda + c_1 R^2 +  c_2 R_{\mu \nu} R^{\mu \nu} + c_3 R_{\mu \nu \rho \sigma } R^{\mu \nu \rho \sigma } \right),
\end{eqnarray}       
where $\Lambda =\frac{(D-1)(D-2)}{2\kappa L^2}$ and $c_i \ll \frac{\alpha'}{L^2}$ \footnote{We represent the dimensions in the bulk by $D$. While the dimensions in the boundary CFT are represented by $d$ ($D=d+1$). }. We are interested in computing the corrections to the shear sum rule in \eqref{egsumrule} to leading order in the coupling constants $c_i$. The main motivation behind such an exercise is the fact that the Hofman Maldacena coefficient $t_2$ is non zero for the class of CFTs that are dual to the quantum corrected classical action. We compute our shear sum rule for the action \eqref{introhdaction} in an $AdS_D$ black hole background by directly computing retarded correlators from gravity and compare it against \eqref{egsumrule}. This serves as a stringent test of the coefficient of $t_2$ in our sum rule \eqref{introdsumrule} which is a consequence of conformal invariance and low energy hydrodynamic behaviour of the theory. 
\paragraph{ } We begin with the analysis of Gauss Bonnet gravity in $D=5$   
\begin{eqnarray}
S_{GB}= \frac{1}{2k_5^2} \int d^5x \sqrt{-g}\left[ R- 2\Lambda + \frac{\lambda_{GB} L^2}{2} (R^2 -4 R_{\mu \nu} R^{\mu \nu} + R_{\mu\nu\rho\sigma}R^{\mu\nu\rho\sigma})\right].
\end{eqnarray}
The sum rule is a consequence of the fact that the retarded correlator is analytic in the upper half plane including the real axis. We show that the retarded correlator is analytic in the upper half plane including the real axis by studying the equations of motion of fluctuations in the bulk that source the stress tensor at the boundary. 
We directly compute the retarded correlator $G_R(\omega)$ in an $AdS_5$ black hole background and examine its high frequency behaviour.
\begin{eqnarray}
\frac{2\epsilon}{3}\left(\frac{8}{5}-\frac{1}{\gamma_{GB}}\right) &=& \lim_{\epsilon \rightarrow 0^+}\frac{1}{\pi}\int_{-\infty}^{\infty} \frac{\delta\rho(z)dz}{z -i\epsilon} ,
\end{eqnarray}
where $\gamma_{GB} = \sqrt{1-4\lambda_{GB}}$. This computation is exact in the Gauss Bonnet coupling $\lambda_{GB}$. We check this against the sum rule calculated using parameters of three point functions and obtain an agreement (see section 5.4 of \cite{Chowdhury:2016hjy}). 
\paragraph{ } We extend this analysis to compute corrections to the sum rule \eqref{egsumrule} due to presence of generic quadratic curvature corrections \eqref{introhdaction}. To leading order in the coupling constants $c_i$s, the modification to the holographic sum rule is given by,
\begin{eqnarray}
\left(\frac{D-1}{2D}+\frac{-4 c_3 (D-4) (D-1) k}{D-2}\right)\epsilon &=& \lim_{\epsilon \rightarrow 0^+}\frac{1}{\pi}\int_{-\infty}^{\infty} \frac{\delta\rho(z)dz}{z -i\epsilon}. 
\end{eqnarray}
For $D=5$, this agrees with the Gauss Bonnet result to leading order in $\lambda_{GB}$. 
In order to compare this against the shear sum rule computed from field theory, we compute the parameters $t_2$ and $t_4$ for possible dual field theories at large $N_c$ but finite 't Hooft coupling $\lambda_c$. Instead of directly computing the three point function of stress tensor from gravity, we compute the same from the conformal collider physics formalism of \citep{Hofman:2009ug, Buchel:2009sk, Myers:2010jv, Sen:2014nfa}. 
\begin{eqnarray}
t_2 = 8c_3\kappa(D-1)(D-2) + O(c_i^2), \qquad t_4 =0.
\end{eqnarray}
Using these parameters of the three point function in \eqref{introdsumrule}, we find agreement with holography to leading order in coupling constants. 
\paragraph{ } The organization of the paper is as follows. We setup the sum rule in section \ref{sumrules}. We compute the sum rule for Gauss Bonnet gravity in section \ref{GBgrav}. We check our holographic result against sum rules obtained from conformal invariance in section \ref{GBfield}. Modifications to the sum rule due to presence of arbitrary higher derivative couplings is discussed in section \ref{gengrav}. In order to compare it against field theory results, we compute the dimensionless parameters $t_2$ and $t_4$ holographically in a conformal collider setup. We obtain a match to leading order in coupling constants $c_i$s.

\section{Sum rules}\label{sumrules}
                       Sum rules result from the analyticity of green's function in the upper half of the complex $\omega$ plane. We are interested in sum rules corresponding to the retarded correlator of the $T_{xy}$ component of the stress tensor in general $d$ dimensions \citep{Chowdhury:2016hjy,David:2011hy,Gulotta:2010cu}. The retarded correlator is defined as 
\begin{eqnarray}
G_R(t,x)=i\theta(t)\langle[T_{xy}(t,x),T_{xy}(0)]\rangle,
\end{eqnarray}
where the expectation value is taken over states at finite temperature and zero chemical potential. The fourier transform is given by,
\begin{eqnarray}\label{ft}
G_R(\omega) &=& \int d^dx e^{i\omega t} G_R(t,x).
\end{eqnarray}
This is consistent with the conventions used in \cite{Chowdhury:2016hjy,David:2011hy,Gulotta:2010cu} . Note that the retarded correlator differs from that used in \citep{Romatschke:2009ng} by a sign. Assuming that the retarded Green's function $G_R(\omega)$ is holomorphic in the upper half plane and has the following convergence property,
\begin{eqnarray}
G_R(\omega) \sim \frac{1}{|\omega|^n} (n > 1), \qquad \textrm{Im } \omega\rightarrow \infty,
\end{eqnarray}
 we can write,
\begin{eqnarray}
G_R(\omega) &=&  \lim_{\epsilon \rightarrow 0^+}\frac{1}{\pi}\int_{-\infty}^{\infty} \frac{\rho(z)dz}{z-\omega -i\epsilon}, 
\end{eqnarray}
where the spectral density is defined as follows,
\begin{eqnarray}
\rho(\omega, p) &=& \textrm{Im }G_R(\omega, p).
\end{eqnarray}
Note that the spectral density function is defined as the imaginary part of the retarded correlator. Usually one finds that the assumption about the well behaved nature of the Green's function is not satisfied. In that case one can consider regularized Green's functions $\delta G_R(\omega)$ such that the second assumption holds true.
\begin{eqnarray}
\delta G_R(\omega) &=&\lim_{\epsilon \rightarrow 0^+}\frac{1}{\pi}\int_{-\infty}^{\infty} \frac{\delta\rho(z)dz}{z-\omega -i\epsilon} ,
\end{eqnarray}
where,
\begin{eqnarray}
\delta\rho(\omega) = \textrm{Im}\left( \delta G_R(\omega) \right).
\end{eqnarray}
The exact regularization procedure depends on the high frequency behaviour. 
\paragraph{ } In order to obtain the high frequency behaviour, let us briefly review the procedure for obtaining retarded correlators from gravity \cite{Son:2002sd}. Given an action (refer to Gb action), we first find black brane solutions to the equations of motion. Such solutions describe a black hole whose horizon is $R^{d-1}$. The boundary theory lives on $R^{d-1,1}$ and the Hawking temperature $T$ of the black hole is identified with the temperature of the dual CFT. Using standard techniques, the thermodynamics of the black hole solution can be studied systematically. In order to obtain the retarded correlator, we consider fluctuations of the black brane metric $\delta g_{\mu \nu}$, which is dual to the stress tensor in the boundary CFT. We solve for the linearized equations of motion for the the fluctuations and impose infalling boundary conditions at the horizon $r^+$ of the black hole. Denoting the radial direction by $r$, we demand the following condition at the boundary.
\begin{eqnarray}
\delta g_{\mu \nu}(r,\omega)|_{r=\infty} = \delta g_{\mu \nu}^0(\omega),
\end{eqnarray}                            
where $\delta g_{\mu \nu}^0(\omega)$ acts as a infinitesimal source for the bulk field $\delta g_{\mu \nu}(r,\omega)$. We plug in this solution into the action, expand to quadratic order in $\delta g_{\mu \nu}(r,\omega)$ and use equations of motion to reduce it to surface terms.
\begin{eqnarray}
S_{OS} &=& -\frac{1}{2}\int \frac{d\omega}{2\pi}  \delta g_{\mu \nu}^0(-\omega) \mathcal{F}(\omega,r)  \delta g_{\mu \nu}^0(\omega)|_{r = \infty},
\end{eqnarray}   
where $S_{OS}$ is the on shell action obtained after integration by parts and imposing equations of motion for fluctuation. The retarded green's function in momentum space is given by
\begin{eqnarray} \label{retgrav}
G_R(\omega)= \mathcal{F}(\omega,r) |_{r=\infty}.
\end{eqnarray} 
\paragraph{ } Usually we find that the large $\omega$ behavior of this quantity is dominated by a divergent piece proportional to $\omega^d f(\omega)$ and a finite contribution leading to breakdown of the assumption of the convergence property of the retarded correlator \citep{ Chowdhury:2016hjy, Gulotta:2010cu, Romatschke:2009ng, David:2012cd}. 
\begin{eqnarray}
\lim_{\omega \rightarrow i\infty} G_R(\omega) \sim \omega^d f\left(\frac{\omega}{\lambda}\right) + {\cal J}.
\end{eqnarray}
We will see that the divergent piece is exactly equal to the fourier transform of the retarded correlator at zero temperature. Following \citep{Romatschke:2009ng, David:2011hy, Chowdhury:2016hjy}, we define the following regularized Green's function
\begin{eqnarray}\label{defgr}
\delta G_R(\omega) &=& G_R(\omega)- G_R(\omega)_{T=0} - {\cal J}.\nonumber\\
\end{eqnarray} 
The sum rule then becomes,
\begin{eqnarray}\label{sumruleexp}
\delta G_R(\omega) &=&\lim_{\epsilon \rightarrow 0^+}\frac{1}{\pi}\int_{-\infty}^{\infty} \frac{\delta\rho(z)dz}{z-\omega -i\epsilon}, 
\end{eqnarray}
where,
\begin{eqnarray}
\delta\rho(\omega) = \rho(\omega)- \rho(\omega)_{T=0}.
\end{eqnarray}
The sum rule given by  \ref{sumruleexp} is evaluated at zero frequency.
\begin{eqnarray}\label{shearsumrule}
\delta G_R(0) &=&\lim_{\epsilon \rightarrow 0^+}\frac{1}{\pi}\int_{-\infty}^{\infty} \frac{\delta\rho(z)dz}{z -i\epsilon} ,
\end{eqnarray}  
where,
\begin{eqnarray}
\delta G_R(0) &=& G_R(0)- G_R(0)_{T=0} - {\cal J}.\nonumber\\
\end{eqnarray}
The low frequency behaviour of the green's function is calculated from relativistic hydrodynamics \citep{Chowdhury:2016hjy}.
\begin{eqnarray}
G_R(0)= P, \qquad G_R(0)_{T=0}=0. 
\end{eqnarray}
The sum rule then becomes,
\begin{eqnarray}
P- \cal{J} &=&\lim_{\epsilon \rightarrow 0^+}\frac{1}{\pi}\int_{-\infty}^{\infty} \frac{\delta\rho(z)dz}{z -i\epsilon} .
\end{eqnarray} 
Note that on the LHS, the first term is from the low frequency hydrodynamic behaviour of the theory while the second term is due to the high frequency behaviour of the retarded correlator. 
%\subsection*{AdS/CFT prescription}

\section{Gauss bonnet gravity in $D=5$} \label{GB}
In this section we evaluate shear sum rules from holography for Gauss Bonnet gravity and match it with field theory calculations. Consider GB gravity defined in $D=5$ dimensions, given by the following action \citep{Myers:2010jv, Buchel:2009sk, Grozdanov:2016fkt}
\begin{eqnarray}\label{GBaction}
S_{GB}= \frac{1}{2k_5^2} \int d^5x \sqrt{-g}\left[ R- 2\Lambda + \frac{\lambda_{GB} L^2}{2} (R^2 -4 R_{\mu \nu} R^{\mu \nu} + R_{\mu\nu\rho\sigma}R^{\mu\nu\rho\sigma})\right].
\end{eqnarray} 
The solutions corresponding to planar AdS black hole solutions are known \cite{Cai:2001dz}.
\begin{eqnarray}\label{GBmetric}
ds^2=  \frac{r^2}{L^2}\left( -\frac{f(r)}{f_\infty} dt^2 + d\bar{x}^2\right) + \frac{L^2}{r^2}\frac{dr^2}{f(r)},
\end{eqnarray} 
where,
\begin{eqnarray}
f(r)&=& \frac{r^2}{L^2} \frac{1}{2\lambda_{GB}} \left[ 1-\sqrt{1-4\lambda_{GB}\left(1- \left(\frac{r^+}{r}\right)^4 \right)} \right] .\nonumber\\
\end{eqnarray}
The horizon for this class of solutions is $r=r^+$. Using the definition,
\begin{eqnarray}
f_\infty &=& \lim_{r \rightarrow \infty} \frac{f(r)}{r^2}=\frac{1-\sqrt{1-4\lambda_{GB}}}{2\lambda_{GB}}=\frac{2}{1+\gamma_{GB}},
\end{eqnarray}
where,
\begin{eqnarray}
\gamma_{GB}=\sqrt{1-4\lambda_{GB}},
\end{eqnarray}
  the coordinates in the metric have been normalized such that the speed of light at the boundary (dual CFT) is one. For $r^+ =0$, we recover the AdS vacuum metric in poincare coordinates with a modified AdS curvature squared scale $\tilde{L}^2=\frac{L^2}{f_\infty}$  \citep{Buchel:2009sk}. The thermodynamics associated with the black brane background  \eqref{GBmetric} is given by \citep{Grozdanov:2016fkt}, 
\begin{eqnarray}\label{thermodyGB}
T &=& \frac{r^+}{\pi \tilde{L}^2} (\frac{1+\gamma_{GB}}{2})^{\frac{3}{2}}, \nonumber\\
s &=& \frac{16 \pi^4 \tilde{L}^3}{k_5^2} \frac{T^3}{(1+\gamma_{GB})^3}, \nonumber\\
\epsilon &=& 3P =\frac{3Ts}{4} ,
\end{eqnarray}  
where $T,s,\epsilon$ and $P$ are the Hawking temperature, entropy, energy density and pressure respectively. We will set $L=1$ for the remainder of this discussion for Gauss Bonnet gravity.

\subsection{Sum rule from gravity}\label{GBgrav}
\paragraph{ } The full gravitational action required for computation of the retarded correlator contains  Gibbons-Hawking surface terms as well as counter terms required for holographic renormalisation. Schematically the action becomes 
\begin{eqnarray}\label{fullaction}
S =S_{GB} + S_{GH} + S_{c.t},
\end{eqnarray}
where $S_{GB}$ is the Gauss Bonnet action given by \ref{GBaction}. The Gibbons Hawking term ($S_{GB}$) and the counter term ($S_{c.t}$) are given in Appendix \ref{GBappendix}.
 In order to compute retarded correlator from this action, we consider fluctuations $\delta g_{xy}$ of the metric in \eqref{GBmetric}, which is dual to stress tensor $T^{xy}$ at the boundary.
 \begin{eqnarray}\label{fluctuationGB}
\delta g_{xy} = \phi(r) e^{-i \omega t}.
\end{eqnarray}  
Introducing the coordinate $u=\left(\frac{r^+}{r}\right)^2$, this obeys the following equation of motion \cite{Grozdanov:2016fkt},
\begin{eqnarray} \label{GBeom}
\partial_u^2 \phi + A(u)\partial_u \phi + B(u)\phi =0, 
\end{eqnarray}
where the terms $A(u)$ and $B(u)$ are given in \eqref{compoeomgb} of appendix \ref{GBappendix}. Following the $AdS/CFT$ prescription outlined in section \ref{sumrules}, the resulting retarded correlator is a sum of three parts \citep{David:2011hy, Chowdhury:2016hjy}.
\begin{eqnarray}\label{retcorGB}
G_R(\omega , T)= \hat{G}_R(\omega, T) + G_{\textrm{counter}}(\omega) + G_{\textrm{contact}}(T),
\end{eqnarray}
where $\hat{G}_R(\omega, T)$ is given by \eqref{retgrav}. $G_{\textrm{counter}}(\omega)$ are counter terms which are required to remove the $\log r$ divergences in $\hat{G}_R(\omega, T)$. They are functions of frequency but not temperature and therefore of the same class as the $T=0$ divergent terms. They cancel when one considers the regularized Green's function $\delta G_R(0)$. The contact term $G_{\textrm{contact}}(T)$ arises from the frequency independent part of the on-shell action. The finite contribution to the regularized Green's function $\delta G_R(0)$ is obtained from the high frequency limit of the $G_R(\omega, T)$.  
\subsubsection*{Analyticity of Green's function in $\omega$ plane}
                   As discussed earlier, we require analyticity of the retarded Green's function $G_R(\omega)$ in the upper half plane for formulation of the sum rules and in this section we justify that assumption. This translates to the fact that the retarded Green's function does not have poles and branch cuts in the upper half plane. The analysis follows closely the arguments outlined in \citep{Gulotta:2010cu,David:2011hy}. We can rewrite the equation of motion of the fluctuation \eqref{GBeom} in terms of the poincare coordinates as ,
\begin{eqnarray}\label{GBeompoincare}
\partial_r^2 \phi(r) + \left( \frac{\partial_r f(r)}{f(r)} + \frac{3}{r} + \frac{\partial_r \alpha(r)}{\alpha(r)}\right)\partial_r \phi + \frac{f_\infty \omega^2}{f(r)^2}\phi =0,
\end{eqnarray}                   
where,
\begin{eqnarray}
f(r)&=&  \frac{r^2}{2\lambda_{GB}} \left[ 1-\sqrt{1-4\lambda_{GB}\left(1- \left(\frac{r^+}{r}\right)^4 \right)} \right], \nonumber\\
f_\infty &=&\frac{1-\sqrt{1-4\lambda_{GB}}}{2\lambda_{GB}},\nonumber\\
\alpha(r) &=& \frac{1}{\sqrt{1-4 \lambda_{GB}\left(1- \frac{(r^+)^4}{r^4}\right)}}.
\end{eqnarray}
Note that $f_\infty$ is positive for $\lambda_{GB} \in [ -\infty , \frac{1}{4} )$. Near the boundary the asymptotics of the functions $f(r)$ and $\alpha(r)$ are given by
\begin{eqnarray}\label{asy1}
\lim_{r \rightarrow \infty}  f(r) &&\sim \frac{\left(1-\sqrt{1-4 \lambda_{GB}}\right) r^2}{2 \lambda_{GB}},\nonumber\\
\lim_{r \rightarrow \infty}  \alpha(r) &&\sim \textrm{constant},
\end{eqnarray}  
while near the horizon we have the following behaviour
\begin{eqnarray}\label{asy2}
\lim_{r \rightarrow r^+}  f(r) &&\sim 4 r^+ (r-r^+),\nonumber\\
\lim_{r \rightarrow r^+}  \alpha(r) &&\sim \textrm{constant}.
\end{eqnarray}  
             The differential equation \eqref{GBeompoincare} can also be obtained by varying the following effective action.
\begin{eqnarray}\label{posactionGB}
S_{\textrm{eff}} &=& \int_{r^+}^{\infty} dr~ r^3f(r)\alpha(r)\left( |\phi'(r)|^2 - \frac{f_{\infty}\omega^2}{f(r)^2}|\phi|^2 \right).
\end{eqnarray}                    
Near the boundary, using \eqref{asy1} the two linearly independent solutions to \eqref{GBeompoincare} are given by
\begin{eqnarray}
\lim_{r \rightarrow \infty}  \phi(r) && \sim \frac{ -\omega ^2\left(1+\sqrt{1-4 \lambda_{GB}}\right)}{4 r^2} I_2\left(\frac{i\omega\sqrt{1+\sqrt{1-4 \lambda_{GB}}}}{\sqrt{2} r}\right), \nonumber\\
&& \sim \frac{ -\omega ^2\left(1+\sqrt{1-4 \lambda_{GB}}\right)}{4 r^2} K_2\left(\frac{i\omega\sqrt{1+\sqrt{1-4 \lambda_{GB}}}}{\sqrt{2} r}\right).
\end{eqnarray}
While near the horizon, using \eqref{asy2}, the leading behaviour is captured by
\begin{eqnarray}
\lim_{r \rightarrow r^+}  \phi(r) \sim (r- r^+)^{\pm\frac{i \omega \sqrt{f_\infty}}{f_0}},
\end{eqnarray}
where
\begin{eqnarray}
\lim_{r \rightarrow r^+} f(r) = f_0 ( r- r^+) + \cdots, \qquad f_0 =4r^+ .
\end{eqnarray}
                    Poles of the retarded Green's function correspond to the existence of quasinormal modes in the spectrum. Quasinormal modes are solutions to the equation \eqref{GBeompoincare} with the following boundary conditions,
\begin{eqnarray}\label{phiqnmasy}
\lim_{r \rightarrow r^+}  \phi(r)&& \sim (r- r^+)^{-\frac{i \omega \sqrt{f_\infty}}{f_0}},\nonumber\\
\lim_{r \rightarrow \infty}  \phi(r)&& \sim r^{-4}.
\end{eqnarray}
We will show systematically that such quasinormal modes cannot exist in the upper half plane for the retarded correlator we are considering. From the time dependence behaviour of the fluctuation \eqref{fluctuationGB}, one can qualitatively see that if quasinormal modes are to exist for $\textrm{Im}~ \omega >0$ (upper half of $\omega$ plane) , these modes will be highly unstable owing to their $e^{-i\omega t}$ dependence. Let us suppose that such quasinormal modes indeed exist with $\textrm{Im}~ \omega >0$. For $\lambda_{GB} \in\left[ -\infty , \frac{1}{4} \right)$, the coefficients in \eqref{GBeompoincare} are real. For a mode $\phi$ with complex frequency $\omega$, we have its complex conjugate $\phi^{*}$ with the frequency $\omega^{*}$. From the action \eqref{posactionGB}, using equations of motion of $\phi$ and $\phi^{*}$, we have
\begin{eqnarray}
0 &=& r^3\alpha(r)f(r)(\phi'^{*}\phi-\phi^*\phi')|_{r^+}^\infty + (\omega^{*2} - \omega^2)\int_{r^+}^{\infty} dr~ \frac{r^3\alpha(r)f_\infty}{f(r)}|\phi|^2.
\end{eqnarray} 
Using the fact that $\textrm{Im}~\omega>0$,  $\lambda_{GB} \in \left[ -\infty , \frac{1}{4} \right)$ and the asymptotic behaviour of $\phi$, $f(r)$ and $\alpha(r)$ from eqns \eqref{phiqnmasy}, \eqref{asy1} and \eqref{asy2}, we find that the above condition is satisfied only when $\omega^2 = \omega^{*2}$. Let us analyse the case when $\omega$ purely imaginary and $\omega > 0$. From \eqref{posactionGB}, we have from equations of motion of $\phi$, 
\begin{eqnarray}
S_{\textrm{eff}} &=& r^3\alpha(r)f(r)\phi'^{*}\phi|_{r^+}^\infty \sim 0.
\end{eqnarray}                 
On the other hand, the action \eqref{posactionGB} is always positive definite for purely imaginary $\omega$. Thus quasinormal modes with purely imaginary $\omega$ do not exist.                    
\paragraph{} For $\omega$ purely real and $\omega \neq 0$, following \citep{David:2011hy, Gulotta:2010cu}, we look at the following wronskian
\begin{eqnarray}
W &=& \phi^*\phi' -\phi'^*\phi,
\end{eqnarray}                   
which satisfies the following equation,
\begin{eqnarray}
W' + \left( \frac{f'(r)}{f(r)} + \frac{3}{r} + \frac{\alpha'(r)}{\alpha}\right) W =0.
\end{eqnarray}                   
From the asymptotic behaviour of the solution \eqref{phiqnmasy} and the functions $f(r)$ and $\alpha(r)$ (eqns \eqref{asy1} and \eqref{asy2}), we examine the quantity $r^3f(r)\alpha(r)W$. When $\omega$ is real, we find that the quasinormal boundary conditions at $ r= \infty$ and $r= r^+$, given by \eqref{phiqnmasy}, cannot be simultaneously satisfied for such a $\phi$. This rules out the possibility of existence of the quasinormal modes for $\textrm{Re}~\omega \neq 0 $. Note that this analysis is valid only for $\lambda_{GB} \in \left[ -\infty , \frac{1}{4} \right)$. 
\paragraph{ } For $\omega=0$, the non existence of quasi normal modes implies that the retarded correlator $G_R(\omega)$ must exhibit an analytic expansion in power series of $\omega$ about the origin, which has been shown explicitly in \cite{Grozdanov:2016fkt}. The authors have computed the small frequency expansion of the retarded correlator in the shear channel for Gauss Bonnet gravity in $D=5$. It has been shown that there are no poles in the retarded Green's function in shear channel for $\omega =0$. More generally, it was shown that for zero spatial momentum and for $\lambda_{GB} \in\left[ -\infty , \frac{1}{4} \right)$, there are no quasi normal modes in the shear channel in the upper half plane \footnote{Note that in \cite{Grozdanov:2016fkt}, the terminology used to describe the shear channel is different. For more details regarding the quasinormal mode spectrum, refer to section 2.1 of \cite{Grozdanov:2016fkt}}.                   Our arguments about analyticity of the Green's function in the upper half plane are consistent with the conclusions derived in \citep{Grozdanov:2016fkt}.
                    
\subsubsection*{High frequency behaviour}
                \paragraph{ } In order to study the high frequency behaviour of the Green's function, we introduce the following variables. 
\begin{eqnarray}\label{changeofvar}
i\lambda = \frac{\omega}{r_+}, \qquad y= \frac{ \lambda r_+}{r}.
\end{eqnarray} 
The equation of motion for $\phi$ becomes,
\begin{eqnarray}\label{diffeqGB1}
\frac{d^2\phi}{dy^2} + \frac{4y^2}{\lambda^3}\left(\frac{\lambda A(y)}{2y} - \frac{\lambda^3}{4y^3}\right)\frac{d\phi}{dy} + \frac{4y^2}{\lambda^4} B(y) =0.
\end{eqnarray}
We expand this equation (\ref{diffeqGB1}) in a power series of $\lambda$
\begin{eqnarray}\label{diffeqGB2}
&&\frac{d^2 \phi}{dy^2}  - \left(\frac{3}{y}+ \frac{2 \left(\left(\gamma_{GB} +1\right) y^3\right)}{\gamma_{GB}^2 \lambda ^4} 
%+ \frac{dy^{2d-1}}{\lambda^2d} 
+ \cdots \right) \frac{d \phi}{dy}\nonumber\\
 && -\left(\frac{\gamma_{GB}^2+2 \sqrt{\gamma_{GB}^2}+1}{2 \gamma_{GB}+2}+\frac{(\gamma_{GB}-1)^3 (\gamma_{GB}+1)^2 y^4}{\left(2 \gamma_{GB}^4+6 \gamma_{GB}^2-2 \sqrt{\gamma_{GB}^2} \left(3 \gamma_{GB}^2+1\right)\right) \lambda ^4}+\cdots \right)\phi=0. \nonumber\\
 \end{eqnarray}
As a check, one can verify that for $\gamma_{GB} =1$ (i.e pure einstein gravity), the leading equation is precisely that of a minimally coupled scalar in pure $AdS_5$ background \citep{David:2011hy}. For non-zero $\gamma_{GB}$, the leading terms in the equation of motion correspond to that of a minimally coupled scalar in a pure $AdS_5$ back ground with the scaled $AdS_5$ curvature $\tilde{L}^2= \frac{L^2}{f_\infty}$. The subleading terms in $\lambda$ correspond to a background of finite temperature, that is the black hole in $AdS_5$. To solve this equation perturbatively in $\frac{1}{\lambda}$, we define
\begin{eqnarray}
\phi = \sum_n \frac{\phi_n(y)}{\lambda^{4n}}.
\end{eqnarray}
We substitute this expansion for $\phi$ in \ref{diffeqGB2} and organise terms order by order in $\frac{1}{\lambda}$,
\begin{eqnarray}
&&\phi_0''(y)-\frac{3}{y}\phi_0'(y)-\frac{1+\gamma_{GB}}{2} \phi_0(y) = 0,\\
&&\phi_1''(y)-\frac{3}{y}\phi_1'(y)-\frac{1+\gamma_{GB}}{2} \phi_1(y)-
\frac{2 \left((\gamma_{GB}+1) y^3\right)}{\gamma_{GB}^2}\phi_0'(y) \nonumber\\
&&\qquad \qquad -\frac{(\gamma_{GB}-1)^3 (\gamma_{GB}+1)^2 y^4}{2 \gamma_{GB}^4+6 \gamma_{GB}^2-2 \left(3 \gamma_{GB}^2+1\right) \gamma_{GB}} \phi_0(y)=0. \nonumber\\
\end{eqnarray}
Note that, as expected the zeroth order solution is the solution to pure $AdS_5$ background with curvature $\tilde{L}^2= \frac{L^2}{f_\infty}$. The two zeroth order solutions are,
\begin{eqnarray}\label{diffeqGB3}
\phi^{(1)}_0 = y^2 K_2\left(\frac{\sqrt{\gamma_{GB}+1} y}{\sqrt{2}}\right), \qquad \phi^{(2)}_0 = y^2 I_2\left(\frac{\sqrt{\gamma_{GB}+1} y}{\sqrt{2}}\right),
\end{eqnarray}
where $K_n$ and $I_n$ are modified bessel functions of second and first kind respectively. We discard the solution $\phi^{(2)}_0$ using the following argument. In the limit $\lambda \rightarrow \infty$, the equations of motion correspond to a minimally coupled scalar in a pure $AdS_5$ background. From the asymptotics of $\phi^{(2)}_0$ we see that the second solution is divergent at the origin $y \rightarrow \infty$ and henceforth not considered for our discussion. Detailed analysis pertaining to this point has been done in \citep{Romatschke:2009ng,David:2012cd}. It can be shown that the second solution does not contribute to the finite term in the high frequency limit. The normalized zeroth order solution is given by,
\begin{eqnarray}\label{solGBzero}
\phi_0=\frac{y^2 K_2\left(\frac{\sqrt{\gamma_{GB}+1} y}{\sqrt{2}}\right)}{\frac{4}{\gamma_{GB}+1}}.
\end{eqnarray}
From \ref{diffeqGB3}, first correction satisfies  
\begin{eqnarray}\label{diffeqGB4}
\phi_1''(y)-\frac{3}{y}\phi_1'(y)-\frac{1+\gamma_{GB}}{2} \phi_1(y)= j(y), \nonumber\\
\end{eqnarray}
where
\begin{eqnarray}
j(y)=\frac{2 \left((\gamma_{GB}+1) y^3\right)}{\gamma_{GB}^2}\phi_0'(y)+\frac{(\gamma_{GB}-1)^3 (\gamma_{GB}+1)^2 y^4}{2 \gamma_{GB}^4+6 \gamma_{GB}^2-2 \left(3 \gamma_{GB}^2+1\right) \gamma_{GB}} \phi_0(y).\nonumber\\
\end{eqnarray}
This is a non-homogeneous second order differential equation, which we solve by Green's function method. We consider the solutions to the homogeneous differential equation
\begin{eqnarray}\label{diffeqGB4hom}
\phi_1''(y)-\frac{3}{y}\phi_1'(y)-\frac{1+\gamma_{GB}}{2} \phi_1(y)=0.
\end{eqnarray}
The two solutions with their respective boundary conditions are given by,
\begin{eqnarray}
f_1(y) = y^2 K_2\left(\frac{\sqrt{\gamma_{GB}+1} y}{\sqrt{2}}\right),\qquad f_1(\infty) =0 \nonumber\\ f_2(y) = y^2 I_2\left(\frac{\sqrt{\gamma_{GB}+1} y}{\sqrt{2}}\right), \qquad f_2(0)=0 .\nonumber\\
\end{eqnarray}
The solution to the full non-homogeneous equation, \ref{diffeqGB4}, is given by
\begin{eqnarray}
\phi_1(y) &=& \int dy' G(y,y') j(y'),
\end{eqnarray}
where the Green's function $G(y,y')$ can be constructed from the two solutions $f_1(y)$ and $f_2(y)$ as follows,
\begin{eqnarray}
G(y,y') &=& -\frac{1}{W(f_1,f_2)(y')}\left( \theta\left(y-y'\right) f_1(y)f_2(y') +\theta\left(y'-y\right) f_1(y')f_2(y) \right).
\end{eqnarray}
The first order correction to pure $AdS_5$ solution is given by
\begin{eqnarray}\label{solGBone}
\phi_1(y) &=& -\left(y^2 K_2\left(\frac{\sqrt{\gamma_{GB}+1} y}{\sqrt{2}}\right)\int_0^y dy' \frac{y'^2 I_2\left(\frac{\sqrt{\gamma_{GB}+1} y'}{\sqrt{2}}\right)j(y')}{y'^3}\right. \nonumber\\
&&\left. + y^2 I_2\left(\frac{\sqrt{\gamma_{GB}+1} y}{\sqrt{2}}\right)\int_y^\infty dy' \frac{y'^2 K_2\left(\frac{\sqrt{\gamma_{GB}+1} y'}{\sqrt{2}}\right)j(y')}{y'^3} \right),
\end{eqnarray}
where
\begin{eqnarray}
j(y')= 
\frac{\left(y'\right)^5 \left(\gamma_{GB} (\gamma_{GB}+1)^3 y' K_2\left(\frac{\sqrt{\gamma_{GB}+1} y'}{\sqrt{2}}\right)-2 \sqrt{2} (\gamma_{GB}+1)^{5/2} K_1\left(\frac{\sqrt{\gamma_{GB}+1} y'}{\sqrt{2}}\right)\right)}{8 \gamma_{GB}^2}. \nonumber\\
\end{eqnarray}
The complete solution is given by,
\begin{eqnarray}\label{solGB}
\phi(y) = \phi_0 (y) + \frac{1}{\lambda^4} \phi_1(y) + \cdots,
\end{eqnarray}
where $\phi_0 (y)$ and $\phi_1(y)$ are given in eqns \eqref{solGBzero} and \eqref{solGBone} respectively and the ellipses denote higher order terms in $\lambda$ which do not contribute to the finite part of our sum rule \citep{David:2012cd}. Let us analyse the behaviour of the solution as $y \rightarrow 0$. Asymptotic expansion of the zeroth order solution is given by,
\begin{eqnarray}\label{asyzeroGB}
\lim_{y \rightarrow 0}\phi_0(y) &=& 1 +O(y^2), \nonumber\\
\lim_{y \rightarrow 0} \phi'_0(y) &=& \frac{1}{4} (-\gamma_{GB}-1) y + O(y^3). \nonumber\\ 
\end{eqnarray} 
For studying the asymptotic behaviour of the first correction to the zeroth order solution  \eqref{solGBone}, we use the following integral identities
\begin{eqnarray}
\int_0^\infty dy y^5 K_2(y)^2 = \frac{32}{5}, \qquad \int_0^\infty dy y^4 K_2(y)K_1(y) = 2. \nonumber\\
\end{eqnarray}
The asymptotic expansion of \eqref{solGBone} is then given by,
\begin{eqnarray}\label{asyoneGB}
\lim_{y \rightarrow 0} \phi_1(y) = -\frac{((\gamma_{GB}+1) (8 \gamma_{GB}-5)) y^4}{20 \gamma_{GB}^2} + O(y^6), \nonumber\\
\lim_{y \rightarrow 0} \phi'_1(y) = -\frac{((\gamma_{GB}+1) (8 \gamma_{GB}-5)) y^3}{5 \gamma_{GB}^2} +O(y^5). 
\end{eqnarray}
\paragraph{ } We are now in a position to evaluate our retarded correlator. Note that higher order corrections in $\lambda$ to the solution \eqref{solGB} do not contribute to the finite term of our sum rule. The on shell action is given by \citep{Grozdanov:2016fkt},
 \begin{eqnarray}\label{onshellGB}
 S_{\textrm{on-shell}} &=& S_1 + S_{\textrm{contact}},\nonumber\\
 \end{eqnarray}
 where,
 \begin{eqnarray}
 S_1 &=& -\lim_{y \rightarrow 0} \frac{\pi^4 T^4}{k_5^2} \int \frac{d\omega}{(2\pi)}\left( -\frac{\sqrt{2}\gamma_{GB} \lambda}{(\gamma_{GB}+1)^{\frac{5}{2}} \left(\frac{y}{\lambda }\right)^3} \phi'(y,-\lambda) \phi(y,\lambda) \right),\nonumber\\
 S_{\textrm{contact}} &=& -PV_4 -\lim_{y \rightarrow 0} \frac{\pi^4 T^4}{k_5^2} \int \frac{d\omega}{(2\pi)} \left( \frac{1}{\sqrt{2}(1+\gamma_{GB})^{\frac{3}{2}}}-\frac{\gamma_{GB} \lambda ^4}{(2(1+\gamma_{GB}))^{\frac{3}{2}} y^2}\right) \phi(y,-\lambda) \phi(y,\lambda).\nonumber\\
  \end{eqnarray}
The retarded correlator and the contact term defined in \eqref{retcorGB} is then given by,
\begin{eqnarray}\label{retGB}
\hat{G}_R(\omega) &=& \lim_{y \rightarrow 0} \frac{2\pi^4 T^4}{k_5^2}\left( -\frac{\sqrt{2}\gamma_{GB} \lambda}{(\gamma_{GB}+1)^{\frac{5}{2}} \left(\frac{y}{\lambda }\right)^3} \phi'(y,-\lambda) \phi(y,\lambda)\right), \nonumber\\
G_{\textrm{contact}} &=& \lim_{y \rightarrow 0} \frac{2\pi^4 T^4}{k_5^2}\left(  \frac{1}{\sqrt{2}(1+\gamma_{GB})^{\frac{3}{2}}}-\frac{\gamma_{GB} \lambda ^4}{(2(1+\gamma_{GB}))^{\frac{3}{2}} y^2}\right) \phi(y,-\lambda) \phi(y,\lambda). \nonumber\\
\end{eqnarray}
Substituting the expansion \eqref{asyoneGB} and \eqref{asyzeroGB} into \eqref{retGB}, we get the high frequency behaviour 
\begin{eqnarray}
\lim_{\lambda \rightarrow \infty} \hat{G}_R(\omega) &=& \left(\lim_{y \rightarrow 0} \frac{-2\pi^4 T^4}{k_5^2} \frac{\sqrt{2}\gamma_{GB} \lambda^4}{(\gamma_{GB}+1)^{\frac{5}{2}}y^3} \phi_0'(y,-\lambda) \phi_0(y,\lambda)\right. 
\nonumber\\
&&\qquad \left. - \frac{(r^+)^4\sqrt{1+\gamma_{GB}}}{\sqrt{2}k_5^2}\left( \frac{8}{5}-\frac{1}{\gamma_{GB}} \right)  + O\left(\frac{1}{\lambda^4}\right)\right),\nonumber\\  
\end{eqnarray}
where we have changed the variables from $r^+$ to $y$ using \eqref{changeofvar}.  Using the asymptotic expansion formulae and black hole thermodynamics, the finite part of the contact term is evaluated to be,
\begin{eqnarray}
G_{\textrm{contact}} &=& \frac{(r^+)^4\sqrt{(1+\gamma_{GB})}}{\sqrt{2}k_5^2} = \frac{\epsilon}{3}.
\end{eqnarray}   
From eqns \eqref{thermodyGB} and \eqref{changeofvar}, we see that the leading term in the high frequency behaviour is proportional to $\omega^4$. As explained before, this divergent term cancels on considering the regularized Green's function $\delta G_R(0)$. We are left with the finite term and the contact term. The contact term cancels with $G_R(0)$ as explained previously. Our sum rule then becomes
\begin{eqnarray}\label{sumruleGBgrav}
\frac{2\epsilon}{3}\left(\frac{8}{5}-\frac{1}{\gamma_{GB}}\right) &=& \lim_{\epsilon \rightarrow 0^+}\frac{1}{\pi}\int_{-\infty}^{\infty} \frac{\delta\rho(z)dz}{z -i\epsilon}. 
\end{eqnarray}
\subsection{Sum rule from field theory}\label{GBfield}
                      \paragraph{} Shear sum rule can also be obtained by studying the thee point function of stress tensor $T_{xy}$. The LHS of the sum rule involves studying the high frequency behaviour of the retarded Green's function of the stress tensor in the shear channel.  For time intervals lesser than the thermal scale ( $\delta t \ll \beta = \frac{1}{T}$), we can analytically continue to the euclidean domain and the operator product expansion (OPE) of the euclidean correlator can be used to study the high frequency behaviour \citep{Romatschke:2009ng, Chowdhury:2016hjy}. 
\begin{eqnarray}\label{OPE}
\langle T_{xy}(s)T_{xy}(0)\rangle \sim C_T\frac{I_{xy,xy}(s)}{s^{8}} + 
\hat{A}_{xyxy\alpha\beta}(s)  \langle T_{\alpha\beta}(0) \rangle + \cdots.
\end{eqnarray}     
A simple dimensional analysis reveals that the fourier transform of the tensor structures scale as follows 
\begin{eqnarray}
 \int  d^4 x e^{i \omega t } C_T \frac{ I_{xy, xy} (x) }{ |s^{8}| }\equiv 
  \sim \omega^4 
 \log( \frac{\omega}{\Lambda} ) , \nonumber\\
 \int d^4 x e^{i\omega t } \hat A_{xyxy\alpha\beta}(x) \langle
T_{\alpha\beta} \rangle \equiv {\cal J}  \sim 
\omega^0 \hat a^{\alpha\beta} \langle T_{\alpha\beta} \rangle.  
\end{eqnarray}
where the expectation value has been taken over states held at finite temperature and zero chemical potential. The divergent part is identical to the one we saw in the previous section, and cancels when one considers the regularized Green's function $\delta G_R(0)$. The high frequency behaviour then consists of the finite piece which is proportional to energy density of the theory. It was shown that the proportionality constant can be written down in terms of Hofman-Maldacena variables $t_2$ and $t_4$ \cite{Hofman:2008ar}, which determine the three point function of stress tensors of a parity preserving CFT \cite{Chowdhury:2016hjy}. The sum rule in $d=4$ becomes,  
\begin{eqnarray}\label{sumruleGBfield}
\frac{\epsilon}{450} (-25 t_2-4 t_4+180) &=& \lim_{\epsilon \rightarrow 0^+}\frac{1}{\pi}\int_{-\infty}^{\infty} \frac{\delta\rho(z)dz}{z -i\epsilon},
\end{eqnarray}   
where
\begin{eqnarray}
t_2=\frac{5 (156 a-4 (9 c-12 b))}{2 (14 a-2 b-5 c)}, \qquad t_4= \frac{15 (4 (5 c-8 b)-81 a)}{2 (14 a-2 b-5 c)},\nonumber\\ 
\end{eqnarray}                   
and $a, b$ and $c$ are the parameters of the three point functions of the stress tensor \cite{Osborn:1993cr}. The collider parameters $t_2$ and $t_4$ are known for Gauss Bonnet gravity theories in $D=5$ \cite{Buchel:2009sk}.
\begin{eqnarray}
t_4 &=& 0, \nonumber\\
t_2 &=& \frac{24 f_\infty \lambda_{GB}}{1-2f_\infty \lambda_{GB}},
\end{eqnarray}
where,
\begin{eqnarray}
f_\infty &=& \frac{1-\sqrt{1-4\lambda_{GB}}}{2\lambda_{GB}} = \frac{2}{1+\gamma_{GB}}.
\end{eqnarray}
Substituting the values of $t_2$ and $t_4$ into the sum rule derived from conformal field theory \eqref{sumruleGBfield}, we get,
\begin{eqnarray}
\frac{2\epsilon}{3}\left(\frac{8}{5}-\frac{1}{\gamma_{GB}}\right) &=& \lim_{\epsilon \rightarrow 0^+}\frac{1}{\pi}\int_{-\infty}^{\infty} \frac{\delta\rho(z)dz}{z -i\epsilon}.
\end{eqnarray}  
This agrees precisely with the holographic calculation \eqref{sumruleGBgrav}. This provides a stringent test of the coefficient of $t_2$ of the sum rule which was derived based on conformal invariance. Note that this agreement also checks the consistency of the effective action of \citep{Grozdanov:2016fkt}. The bounds on the collier parameters $t_2$ and $t_4$ are known  \cite{Buchel:2009sk}. This constrains the Gauss bonnet coupling to lie between the following bounds
\begin{eqnarray}
-\frac{(3d+2)(d-2)}{4(d+2)^2} \leq \lambda_{GB} \leq \frac{(d-2)(d-3)(d^2-d+6)}{4(d^2-3d+6)^2}
\end{eqnarray} 
This translates to constraints on the shear sum rule
\begin{eqnarray}
\left( \frac{1}{2} + \frac{1}{d+1} \right ) P\leq \delta G_{GB}(0) \leq 
\left( \frac{d}{2} - \frac{d-1}{2(d+1)} \right) P
\end{eqnarray}
It is easy to see that this is within the general bound derived for the sum rule 
in \cite{Chowdhury:2016hjy}.                  
                    
\section{General higher derivative corrections in $D \geq 5$} \label{gen}
                                      We generalize our observations about the shear sum rule for theories with generic higher derivative corrections, truncated at fourth order in derivatives, in arbitrary dimensions. The curvature squared corrections to pure Einstein-Hilbert action is given by \citep{Kats:2007mq}
\begin{eqnarray}\label{actionGEN}
S=\int d^D x \sqrt{-g} \left( \frac{R}{2\kappa}- \Lambda + c_1 R^2 +  c_2 R_{\mu \nu} R^{\mu \nu} + c_3 R_{\mu \nu \rho \sigma } R^{\mu \nu \rho \sigma } \right), 
\end{eqnarray}
where $\Lambda =\frac{(D-1)(D-2)}{2\kappa L^2}$. Following \cite{Kats:2007mq}, we consider perturbative corrections to the pure Einstein gravity black hole solutions due to the presence of curvature squared terms. Consider the Einstein-Hilbert action in general $D$ dimensions.
\begin{eqnarray}\label{Einstein gravity}
S=\int d^D x \sqrt{-g} \left( \frac{R}{2\kappa}- \Lambda \right).
\end{eqnarray}  
We can write down the black brane solutions to this action. Denoting $u=\left(\frac{r_0}{r} \right)^{\frac{\left( D-1 \right)}{2}}$ the metric that solves the equations of motion of the general Einstein-Hilbert action is given by,
\begin{eqnarray}\label{genbhsol}
ds^2=r_0^2\frac{-f(u)dt^2 + d\bar{x}^2}{L^2u^{\frac{4}{(D-1)}}} + \frac{4L^2}{(D-1)^2}\frac{du^2}{u^2 f(u)},\qquad f(u)=1-u^2.
\end{eqnarray} 
The effect of the higher derivative corrections in \eqref{actionGEN} is to modify the function $f(u)$\footnote{Henceforth all our computations for the higher derivative corrections are valid upto leading order in $c_1$, $c_2$ and $c_3$.}.
\begin{eqnarray}\label{genfdef}
f(u)&=&1-u^2 +\alpha + \gamma u^4, \nonumber\\
 \end{eqnarray}
where,
\begin{eqnarray}
\alpha=\frac{2 (D-4) \kappa ((D-1) (D c_1 +c_2)+2 c_3)}{L^2(D-2)}, \qquad \gamma= \frac{2 c_3 (D-3) (D-4) \kappa}{L^2},
\end{eqnarray}
As with the Gauss Bonnet case, we assume $L=1$ for remainder of the discussion.
In these coordinates, the horizon of the black hole to leading order is given by
\begin{eqnarray}
u_H &=& 1+\frac{(\alpha + \gamma)}{2},
\end{eqnarray}
In terms of poincare coordinates,
\begin{eqnarray}\label{genhorizon}
r_H &=&  r_0\left(1-\frac{(\alpha + \gamma)}{D-1}\right).
\end{eqnarray}
The thermodynamics of the black hole solution is given by \citep{Kats:2007mq}
\begin{eqnarray}\label{thermodygen}
T &=& \frac{(D-1) r_H }{4 \pi }\left(\frac{(D-4) \kappa ((D-1) (c_1 D+c_2)+2 c_3)}{D-2}-2 c_3 (D-4) (D-3) k+1\right), \nonumber\\
s &=& \frac{2 \pi r_H^{D-2}}{k}\left(-4 c_1 \kappa (D-1) D -4 c_2 \kappa(D-1) + 4  c_3 \kappa (D-4) (D-1)+1  \right), \nonumber\\
\epsilon &=& \frac{r_H^{D-1}}{2 k}\left(\kappa (c_1 D ((7-3 D) D-4)+ c_2 ((7-3 D) D-4)+2 c_3 (D ((D-5) D+3)+4))+D-2\right). \nonumber\\
\end{eqnarray} 
Note that the coordinate $t$ is not the natural candidate for time in the boundary theory. Following \citep{Kats:2007mq}, as $u \rightarrow 0$,
\begin{eqnarray}
ds^2 \sim -f(0)dt^2 + d\bar{x}^2.
\end{eqnarray} 
The boundary time coordinate is then $t_b = \sqrt{f(0)} t$. The fourier transform in the bulk is then defined as,
\begin{eqnarray}
G_R(\omega) &=& \int dt_b \phantom{r} e^{i \omega t_b} G_R(t).
\end{eqnarray}

\subsection{Sum rule from gravity} \label{gengrav}
                \paragraph{ } The shear sum rule for pure Einstein gravity (\eqref{Einstein gravity}) in general dimensions $D=d+1$ is given by \citep{Gulotta:2010cu, Chowdhury:2016hjy},
\begin{eqnarray}\label{egsumrule2}
\frac{d \epsilon }{2(d+1)} &=& \lim_{\epsilon \rightarrow 0^+}\frac{1}{\pi}\int_{-\infty}^{\infty} \frac{\delta\rho(z)dz}{z -i\epsilon}.  
\end{eqnarray} 
We will now evaluate corrections to \eqref{egsumrule2} due to curvature squared corrections to Eistein-Hilbert action.
 
\paragraph{ } The retarded correlator is given by,
\begin{eqnarray}\label{retcorgen}
G_R(\omega , T)= \tilde{G}_R(\omega, T) + \tilde{G}_{\textrm{counter}}(\omega) + \tilde{G}_{\textrm{contact}}(T),
\end{eqnarray}
where as before, $\tilde{G}_R(\omega, T)$ is the retarded correlator from gravity as defined in \eqref{retgrav} and  $\tilde{G}_{\textrm{counter}}(\omega)$ is the counter term which cancels on considering the regularized Green's function. The finite part of the sum rule is controlled by the high frequency behaviour of the retarded correlator. We consider fluctuations of the form,
\begin{eqnarray}
\delta g_{xy} = \tilde{\phi}(r) e^{-i \omega t_b}.
\end{eqnarray}
The fluctuation obeys the following equation of motion
\begin{eqnarray}\label{geneom}
\tilde{\phi}''(u) + C(u) \tilde{\phi}'(u) + D(u) \tilde{\phi}(u) =0 .
\end{eqnarray}
where, 
$C(u)$ and $D(u)$ are given in \eqref{compoeomgen} of Appendix \ref{genappendix}. 
In order to study the high frequency behaviour, we introduce the variables  
\begin{eqnarray}
i\lambda = \frac{\omega}{r_H}, \qquad y= \frac{ \lambda r_H}{r},
\end{eqnarray}
where $r_H$ is the horizon of the black hole with the curvature square corrections. It is related to the horizon radius $r_0$ of the pure Einstein-Hilbert action by \eqref{genhorizon}. In these coordinates the equation for $\tilde{\phi}$ becomes,
\begin{eqnarray}
\tilde{\phi}''(y) + \left( \frac{C(y)(D-1) \left(\frac{\alpha +\gamma }{D-1}+1\right)^{\frac{D-1}{2}}}{2 \lambda ^{\frac{D-1}{2}} y^{\frac{3-D}{2}}}+\frac{3-D}{2 y} \right) \tilde{\phi}'(y) - \frac{f(0)}{f(y)^2} \tilde{\phi}(y) =0.
\end{eqnarray} 
We are interested in solutions to this equation to linear order in the coupling constants $c_1, c_2$ and $c_3$. To leading order in the coupling constants, we expand this in a power series of $\lambda$.
\begin{eqnarray}\label{genpert1}
&&\tilde{\phi}''(y) + \left( \frac{2-D}{y} -\left((D-1) y^{D-2} +\frac{ \gamma  \left(D^2-1\right) y^{D-2}}{(D-3)}\right)\frac{1}{ \lambda ^{D-1}} + \cdots \right)\tilde{\phi}'(y) \nonumber\\
&&\qquad \qquad + \left( (\alpha -1)+\left( \left(2 \alpha -2 \gamma \right) -2 \right)\frac{y^{D-1}}{\lambda ^{D-1}}+ \cdots \right) \phi(y) =0.
\end{eqnarray}
where the ellipses denote terms of higher order in $\lambda$. In the strict $\lambda \rightarrow \infty$ limit, we recover minimally coupled scalar in a pure $AdS_D$ background with curvature scale $\tilde{L}^2 = \frac{1}{(1+\alpha)}$. We assume the follwing ansatz for $\tilde{\phi}$ to solve \eqref{genpert1}, order by order in $\frac{1}{\lambda}$.
\begin{eqnarray}
\tilde{\phi} = \sum_n \frac{\tilde{\phi}_n(y)}{\lambda^{n(D-1)}}.
\end{eqnarray}   
The zeroth order solution and the first correction to the zeroth order obey the following equations of motion.
\begin{eqnarray}\label{genpert2}
&&\tilde{\phi}_0''(y)+\frac{2-D}{y}\tilde{\phi}_0'(y)+(-1 + \alpha)\tilde{\phi}_0(y) = 0,\\
&& \tilde{\phi}_1''(y)+\frac{2-D}{y}\tilde{\phi}_1'(y)+(-1 + \alpha)\tilde{\phi}_1(y) = \tilde{j}(y),
\end{eqnarray} 
where
\begin{eqnarray}
\tilde{j}(y)=\left(\frac{\gamma  \left(D^2-1\right) y^{D-2}}{D-3}+(D-1) y^{D-2}\right) \tilde{\phi}_0'(y)+ \left(-2 \alpha  y^{D-1}+2 \gamma  y^{D-1}+2 y^{D-1}\right) \tilde{\phi}_0(y).\nonumber\\
\end{eqnarray}
The normalized zeroth order solution to leading order in the coupling constants is given by 
\begin{eqnarray}
\tilde{\phi}_0(y) &=& \frac{2^{\frac{3}{2}-\frac{D}{2}} y^{\frac{D-1}{2}} K_{\frac{D-1}{2}}(y)}{\Gamma \left(\frac{D-1}{2}\right)} + \alpha \frac{ 2^{\frac{1}{2}-\frac{D}{2}} y^{\frac{D+1}{2}} K_{\frac{D-3}{2}}(y)}{\Gamma \left(\frac{D-1}{2}\right)},
\end{eqnarray}
where as before we consider only the solution that remains finite as $y \rightarrow \infty$.
The first correction to the zeroth order solution is obtained by solving \eqref{genpert2}.
\begin{eqnarray}
\tilde{\phi}_1''(y)+\frac{2-D}{y}\tilde{\phi}_1'(y)+(-1 + \alpha)\tilde{\phi}_1(y) = \tilde{j}(y),
\end{eqnarray}
where,
\begin{eqnarray}
\tilde{j}(y)&=& \gamma \frac{ 2^{\frac{1}{2}-\frac{D}{2}} y^{\frac{3 D}{2}-\frac{5}{2}} \left(\left(2-2 D^2\right) K_{\frac{D-3}{2}}(y)+(4 D-12) y K_{\frac{D-1}{2}}(y)\right)}{(D-3) \Gamma \left(\frac{D-1}{2}\right)} \nonumber\\
&&+\frac{2^{\frac{1}{2}-\frac{D}{2}} y^{\frac{3 D}{2}-\frac{5}{2}} \left(\left(\alpha +\alpha  D^2-2 (\alpha +1) D+2 \alpha  y^2+2\right) K_{\frac{D-3}{2}}(y)-y (\alpha  (D+3)-4) K_{\frac{D-1}{2}}(y)\right)}{\Gamma \left(\frac{D-1}{2}\right)}. \nonumber\\
\end{eqnarray}
Solution to this non-homogenous equation is obtained by Green's function method. 
Using the solutions of the homogenous equation
\begin{eqnarray}
\tilde{\phi}_1''(y)+\frac{2-D}{y}\tilde{\phi}_1'(y)+(-1 + \alpha)\tilde{\phi}_1(y) =0,
\end{eqnarray}
we can construct a Green's function,
\begin{eqnarray}
\tilde{G}(y,y') &=& -\frac{1}{W(\tilde{f}_1,\tilde{f}_2)(y')}\left( \theta\left(y-y'\right) \tilde{f}_1(y)\tilde{f}_2(y') +\theta\left(y'-y\right) \tilde{f}_1(y')\tilde{f}_2(y) \right), \nonumber\\
\end{eqnarray}
where
\begin{eqnarray}
\tilde{f}_1(y) &=& y^{\frac{D-1}{2}} K_{\frac{D-1}{2}}(y) +\frac{1}{4} \alpha  y^{\frac{D-1}{2}} \left(2 y K_{\frac{D+1}{2}}(y)-(D-1) K_{\frac{D-1}{2}}(y)\right) \nonumber\\
\tilde{f}_2(y) &=&y^{\frac{D-1}{2}} I_{\frac{D-1}{2}}(y)+\frac{1}{4} \alpha  y^{\frac{D-1}{2}} \left(-2 y I_{\frac{D+1}{2}}(y)-(D-1) I_{\frac{D-1}{2}}(y)\right).
\end{eqnarray}
are the two solutions of the homogeneous equation.
The first correction is then given by,
\begin{eqnarray}
\tilde{\phi}_1(y) &=& \int dy' \phantom{d} \tilde{G}(y,y') \tilde{j}(y').  
\end{eqnarray}
The complete solution is then given by
\begin{eqnarray}\label{solgen}
\tilde{\phi}(y) = \tilde{\phi}_0 (y) + \frac{1}{\lambda^{D-1}} \tilde{\phi}_1(y) + \cdots.
\end{eqnarray}
Let us look at the asymptotic behaviour of the solution.
\begin{eqnarray}\label{asyzerogen}
\lim_{y \rightarrow 0}\tilde{\phi}_0(y) &=& 1 +O(y^2), \nonumber\\
\lim_{y \rightarrow 0} \tilde{\phi}'_0(y) &=& \frac{(\alpha -1) y}{D-3} + O(y^3), \nonumber\\ 
\lim_{y \rightarrow 0} \tilde{\phi}''_0(y) &=& \frac{\alpha -1}{D-3} + O(y^2). \nonumber\\
\end{eqnarray}
Using integrals listed in \eqref{integrals} of Appendix \ref{genappendix}, we obtain the asymptotic behaviour of the first order correction $\tilde{\phi}_1(y)$. 
\begin{eqnarray}\label{asyonegen}
\lim_{y \rightarrow 0}\tilde{\phi}_1(y) &=& -\frac{6 (\gamma +1)+D (\gamma  (D-9)+D-5)}{2 ((D-3) D) } y^{D-1} + \cdots ,\nonumber\\
\lim_{y \rightarrow 0}\tilde{\phi}'_1(y) &=& -\frac{(D-1) (6 (\gamma +1)+D (\gamma  (D-9)+D-5))}{2 ((D-3) D)}y^{D-2} + \cdots ,\nonumber\\
\lim_{y \rightarrow 0}\tilde{\phi}''_1(y) &=& -\frac{(D-2) (D-1) (6 (\gamma +1)+D (\gamma  (D-9)+D-5))}{2 ((D-3) D)} y^{D-3} + \cdots .\nonumber\\  
\end{eqnarray}
The on shell action with the relevant Gibbons-Hawking boundary terms as well as the counter terms is given by \citep{Kats:2007mq}.
\begin{eqnarray}
\tilde{S}_{\textrm{On shell}} = \left(-\frac{V_{D-2}r_H^{D-1} (1+\frac{\alpha + \gamma}{D-1})^{D-1}}{16\kappa \sqrt{f(0)}} \int \frac{d\omega}{2\pi} \phantom{d}\lim_{y\rightarrow 0} \mathcal{I}\right) + \tilde{S}_{\textrm{contact}},
\end{eqnarray} 
where \footnote{The action in \eqref{onshellgen} has an overall negative sign compared to \citep{Kats:2007mq}. Our convention for the retarded correlator from the on shell action differs from \citep{Kats:2007mq} by a negative sign.We define $G_R= -\frac{\delta^2 S_{\textrm{On-shell}}}{\delta \phi^2}$ as in \citep{Son:2002sd}. We have verified that with this extra negative sign, \eqref{onshellgen} agrees with the on shell action \eqref{onshellGB} to first order in $\lambda_{GB}$.}
\begin{eqnarray}\label{onshellgen}
\mathcal{I} &=&  \left( \frac{2 \lambda ^{\frac{D-1}{2}} y^{\frac{3-D}{2}} }{(D-1) \left(\frac{\alpha +\gamma }{D-1}+1\right)^{\frac{D-1}{2}}}\left(-\tilde{A}(y) + \tilde{B}(y)-\frac{\left(2 \lambda ^{\frac{D-1}{2}} y^{\frac{3-D}{2}}\right) \tilde{F}'(y)}{2 \left((D-1) \left(\frac{\alpha +\gamma }{D-1}+1\right)^{\frac{D-1}{2}}\right)}\right)\tilde{\phi}'(-\omega)\tilde{\phi}(\omega)\right) \nonumber\\
&& -\frac{\left(4 \lambda ^{D-1} y^{\frac{3-D}{2}}\right) \left(\left(2 \lambda ^{\frac{D-1}{2}} y^{\frac{3-D}{2}}\right) \tilde{E}'(y)\right)}{\left((D-1) \left(\frac{\alpha +\gamma }{D-1}+1\right)^{\frac{D-1}{2}}\right) \left((D-1)^2 \left(\frac{\alpha +\gamma }{D-1}+1\right)^{D-1}\right)}\nonumber\\
&&\times\tilde{\phi}(y) \left(\frac{1}{2} (3-D) y^{\frac{1-D}{2}} \tilde{\phi}'(y)+y^{\frac{3-D}{2}}  \tilde{\phi}''(y)\right)+\cdots,
\end{eqnarray}
where $A,B,C$ and $E$ are given in \eqref{compoeffgen} of appendix \ref{genappendix}. The ellipses denotes terms in the on-shell action which donot contribute to the finite part of the sum rule. The details of the contact term $\tilde{S}_{\textrm{contact}}$ is not required. Upon taking the limit $y \rightarrow 0$, this contributes a constant pressure term to the sum rule (analogous to the Gauss Bonnet case). The retarded correlator defined in \ref{retcorgen} then takes the following form
\begin{eqnarray}
\tilde{G}_R(\omega, T) &=& \lim_{y \rightarrow 0}\frac{r_H^{D-1} (1+\frac{\alpha + \gamma}{D-1})^{D-1}}{8\kappa \sqrt{f(0)}}\phantom{d}\mathcal{I}(y) ,\nonumber\\
\tilde{G}_{\textrm{contact}}(T) &=& P.
\end{eqnarray}
Using the asymptotic behaviour of solutions in \eqref{asyzerogen} and \eqref{asyonegen}, the high frequency behaviour of the retarded correlator is obtained. The finite part of the high frequency behaviour is given by
\begin{eqnarray}
\mathcal{J} &=& -\frac{(\frac{D}{2}+\frac{1}{D}-\frac{3}{2}-4 c_3 (D-4) (D-1) \kappa)\epsilon}{D-2}+\frac{\epsilon}{D-2} .
\end{eqnarray}
where the second contribution is due to the contact term and the energy density $\epsilon$ is defined in \eqref{thermodygen}. The regularized sum rule at zero frequency becomes,
\begin{eqnarray}
\left(\frac{D-1}{2D}+\frac{-4 c_3 (D-4) (D-1) \kappa}{D-2}\right)\epsilon &=& \lim_{\epsilon \rightarrow 0^+}\frac{1}{\pi}\int_{-\infty}^{\infty} \frac{\delta\rho(z)dz}{z -i\epsilon}. 
\end{eqnarray}
With the identification $D=d+1$,
\begin{eqnarray}\label{gensumruleholo}
\left(\frac{d}{2(d+1)}+\frac{-4 c_3 d(d-3) \kappa}{d-1}\right)\epsilon &=& \lim_{\epsilon \rightarrow 0^+}\frac{1}{\pi}\int_{-\infty}^{\infty} \frac{\delta\rho(z)dz}{z -i\epsilon}.
\end{eqnarray}
\subsection{Sum rule from field theory}\label{genfield}
                       \paragraph{ } Shear sum rule for a general conformal field theory in arbitrary dimensions is known. LHS of the sum rule is a function of  $a,b$ and $c$, which are interpreted as the parameters of the three point function of the stress tensor in the $D-1$ dimensional conformal field theory \cite{Osborn:1993cr}, which is the possible dual of the quantum corrected action \eqref{actionGEN}.
\begin{eqnarray}\label{gensumrule}
\left( \frac{D-1}{2D} + \frac{(-D+4) t_2}{2 (D-2)^2} + \frac{\left(-D^2+5 D-2\right) t_4}{(D-2)^2 D^2} \right) \epsilon &=& \lim_{\epsilon \rightarrow 0^+}\frac{1}{\pi}\int_{-\infty}^{\infty} \frac{\delta\rho(z)dz}{z -i\epsilon},\nonumber\\
\end{eqnarray}                                
where 
\begin{eqnarray}\label{t2t4}
t_2 &=& \frac{2 D (a (D-2) (D (D+6)-3)+(D-1) (3 b (D-1)-2 c D+c))}{(D-1) (a (D-3) (D+2)-2 b-c D)}, \nonumber\\
t_4 &=& -\frac{D (D+1) (3 a (D-2) (2 D-1)+(D-1) (2 b (D-1)-c D))}{(D-1) (a (D-3) (D+2)-2 b-c D)}.
\end{eqnarray}
In order to compare our holographic sum rule with field theory computation, we need to calculate the parameters $t_2$ and $t_4$ holographically to leading order in the coupling constants. 
\paragraph{}  Direct holographic computation of three point function using the methods of \citep{Arutyunov:1999nw} is a tedious task. Instead, we compute the dimensionless parameters $t_2$ and $t_4$ from the conformal collider physics formalism of \cite{Hofman:2008ar, Hofman:2009ug, Buchel:2009sk, Myers:2010jv}\footnote{For CFTS in $d=3$ such conformal collider bounds have been studied in \cite{Chowdhury:2017vel,Cordova:2017zej}.}. Consider localised perturbations of the $d$ dimensional minkowski CFT at the origin.
The perturbations evolve and spread out in time. We measure the integrated energy flux per unit angle over states created by such perturbations, at a large sphere of radius $r$. 
\begin{eqnarray}\label{energyfunc}
\langle {E}_{\hat n} \rangle &=& \frac{\langle 0|\mathcal{O}^\dagger{E}_{\hat n}
\mathcal{O}|0\rangle}{\langle 0|\mathcal{O}^\dagger \mathcal{O}|0\rangle},\nonumber\\                    
{E}_{\hat n}  &=& \lim_{r \rightarrow \infty} r^{(d-2)} \int_{-\infty}^{\infty} dt n^iT^{t}_i(t, r\hat{n}),\nonumber\\
\mathcal{O} &\sim & \frac{\epsilon^{ij} T_{ij}}{\sqrt{\langle \epsilon^{ij}T_{ij}|T_{ij}\epsilon^{ij}\rangle}}, \qquad \frac{\epsilon^{i} j_{i}}{\sqrt{\langle \epsilon^{j}j_{j}|j_{i}\epsilon^{i}\rangle}},\nonumber\\
\end{eqnarray}   
\\                    
where, $\mathcal{O}$ is the operator creating the perturbation and  $\hat n$ is a unit vector on $R^{d-1}$, which specifies the position of the 
calorimeter on the sphere. For perturbations created by stress tensor insertions, $\mathcal{O} \sim \epsilon_{ij} T^{ij}$, the energy flux (for $d > 3$) takes the following form, 
\begin{eqnarray}\label{energyfngenform}
\langle {E}_{\hat n} \rangle &=& \frac{E}{\Omega_{d-2}}\left( 1+ t_2\left(\frac{\epsilon_{\alpha \beta} \epsilon^*_{\alpha \gamma} n^{\beta} n^{\gamma}}{\epsilon_{\alpha \beta}\epsilon^*_{\alpha \beta}}- \frac{1}{d-1}\right) + t_4\left( \frac{|\epsilon_{\alpha \beta}n^{\alpha}n^{\beta}|^2}{\epsilon_{\alpha \beta}\epsilon^*_{\alpha \beta}}-\frac{2}{d^2-1}\right)\right),\nonumber\\
\end{eqnarray}
where $E$ is the energy flux and $\Omega_d = \frac{2\pi^{\frac{d+1}{2}}}{\Gamma(\frac{d+1}{2})}$ is the area of a unit $d$-sphere. The dimensionless parameters $t_2$ and $t_4$ is a function of parameters of the three point function of stress tensor and has been defined in \eqref{t2t4}.

\subsubsection*{Normalization of two point function}
                     In order to compute the energy functional we will also need the normalisation of the two point function of the stress tensor. In $d$ dimensional CFT the two point function takes the following form \citep{Osborn:1993cr},
\begin{eqnarray}\label{2ptfncft}
\langle T_{\mu \nu}(s) T_{\rho \sigma}\rangle &=& \frac{\mathcal{C}_T}{s^{2d}}\mathcal{I}_{\mu\nu, \rho\sigma}, \nonumber\\
\end{eqnarray}                     
where,
\begin{eqnarray}
\mathcal{E}^T_{\mu\nu, \alpha \beta} &=& \frac{1}{2}(\eta_{\mu \alpha} \eta_{\nu \beta} + \eta_{\mu \beta} \eta_{\nu \alpha})- \frac{1}{d} \eta_{\alpha \beta} \eta_{\mu \nu},\\
I_{\alpha \beta} (x) &=& \eta_{\alpha \beta} -\frac{2 x_\alpha x_\beta}{x^2},\nonumber\\
\mathcal{I}_{\mu \nu, \alpha \beta}(x) &=& I_{\mu \mu'} (x) I_{\nu \nu'} (x) {\mathcal{E}^{T}}^{\mu' \nu',}_{\phantom{\mu' \nu'}\alpha \beta},\nonumber\\
\end{eqnarray}                     
and $\mathcal{C}_T$ is the central charge. We compute $\mathcal{C}_T$ holographically by looking at the specific case $\langle T_{xy} T_{xy}\rangle$. It is sufficient to compute this two point function in pure $AdS_D$ background. We turn on perturbations $h_{xy} = \phi(r) e^{i \omega t_b}$ in \eqref{genbhsol}, with $f(u)$ given by \eqref{genfdef}, after setting $r_H=0$. The equation of motion for $\phi$ becomes,
\begin{eqnarray}
\phi''(r)+\frac{D\phi'(r)}{r}-\frac{\omega^2 \phi(r)}{f(0)r^4}=0.
\end{eqnarray}
The two solutions to this equation is given by,
\begin{eqnarray}
\phi_1(r) &=& C_1 \frac{\omega ^{\frac{D-1}{2}} K_{\frac{D-1}{2}}\left(\frac{\omega }{r \sqrt{f(0)}}\right)}{r^{\frac{D-1}{2}}}, \qquad \phi_2(r)=C_2\frac{\omega ^{\frac{D-1}{2}} I_{\frac{D-1}{2}}\left(\frac{\omega }{r \sqrt{f(0)}}\right)}{r^{\frac{D-1}{2}}},  
\end{eqnarray}
where $C_1$ and $C_2$ are arbitrary constants. We set $C_1=\frac{1}{2^{\frac{D-3}{2}} (f(0))^{\frac{D-1}{4}} \Gamma \left(\frac{D-1}{2}\right)}$ and $C_2=0$ in order to satisfy the boundary condition that the solutions remain finite at $r=0$. Using equations of motion, the action \eqref{actionGEN} reduces to a quadratic function of $\phi$ and its derivatives in fourier space.
\begin{eqnarray}\label{gen2ptfnaction}
S &=& -V_{D-2} \int \frac{d\omega}{2\pi} \tilde{C}(r)\phi(r,-\omega)\phi'(r,\omega)+ \tilde{D}(r)\phi (r, -\omega) \left((D+1) \phi '(r ,\omega)+2 r \phi ''(r, \omega)\right)\nonumber\\
&&\phantom{-V_{D-2} \int}+ \cdots, \nonumber\\
\end{eqnarray} 
where $\tilde{C}(r)$ and $\tilde{D}(r)$ are given in \eqref{coeffaction2ptfn}. The ellipses represent 
terms in the on shell action which won't contribute to the two point function. Using the following series expansions,
\begin{eqnarray}
K_n(z) &=& \frac{1}{2}(\frac{1}{2}z)^{-n}\sum_{k=0}^{n-1} \frac{\left( n-k-1 \right)!}{k!}\left(-\frac{1}{4}z^k\right)\nonumber\\
&&+(-1)^{n+1}\log\left(\frac{z}{2}\right) I_n(z)\nonumber\\
&&+(-1)^n\frac{1}{2}\left(\frac{z}{2}\right)^n\sum_{k=0}^\infty \left( \psi(k+1)+\psi(k+n+1)\right)\frac{\left( \frac{z^2}{4}\right)^k}{k!(n+k!)}, \nonumber\\
I_n(z)&=&(\frac{1}{2}z)^{n} \sum_{k=0}^\infty  \frac{\left(\frac{z^2}{4}\right)^k}{k!\Gamma(n+k+1)},
\end{eqnarray} 
we extract terms proportional to $\log |\omega|$ in the two point function. %\footnote{There is a discrepancy of sign in the ft obtained from gravity calculations and that obtained from ft of field theory correlator. this is solved by the fact that gravity result gives euclidean FT. so in field theory side we have to do it in euclidean signature. also this is not related to prescription of retarded green's function}. 
We get
\begin{eqnarray}
\langle T_{xy} T_{xy} \rangle \sim \frac{2^{2-D}  \omega ^{D-1} \log (\omega ) (1-4 \kappa (c_1 (D-1) D+c_2 (D-1)-2 c_3 (D-4)))}{\kappa f(0)^{\frac{D}{2}-1}\Gamma \left(\frac{D-1}{2}\right)^2} \nonumber\\
\end{eqnarray} 
We compare this against the $d$ dimensional CFT result. Fourier transforming the CFT two point function \eqref{2ptfncft}, we have \cite{Sen:2014nfa, Buchel:2009sk,Myers:2010jv},
\begin{eqnarray}
G_R(\omega) &=& \frac{(d-1) \pi ^{d/2}\mathcal{C}_T}{(d+1) \left(2^{d-1} \Gamma (d+1) \Gamma \left(\frac{d}{2}\right)\right)} \omega^{d} \log \omega + G_{\textrm{analytic}}
\end{eqnarray} 
where $D=d+1$ and $G_{\textrm{analytic}}$ denotes terms wich are analytic in $\omega$. Comparing we have,
\begin{eqnarray}
\mathcal{C}_T &=&   \tilde{\alpha}(1-4 \kappa (c_1 (D-1) D+c_2 (D-1)-2 c_3 (D-4))), \nonumber\\
\tilde{\alpha} &=& \frac{\pi ^{\frac{1}{2}-\frac{D}{2}} \Gamma (D+1)}{(D-2) f(0)^{\frac{D}{2}-1}\kappa \Gamma \left(\frac{D-1}{2}\right)}.
\end{eqnarray} 
\subsubsection*{Holographic computation of $t_2$}                                    
 \paragraph{} We are now in a position to compute the energy functional. By construction, the energy functional for stress tensor perturbations is a function of three point function, normalized by the two point function of the stress tensor. We consider the following coordinate transformation,
\begin{eqnarray}\label{coord}
y^+=-\frac{1}{x^+}, \qquad y^-=x^- -\frac{x_\alpha x^\alpha}{x^+},\qquad y^\alpha = \frac{x^\alpha}{x^+}, \nonumber\\
\end{eqnarray}   
where, $x^\pm = x^0 \pm x^{d-1}$. In these coordinates, the energy flux is measured at $y^+ =0$. Moreover the positivity of energy flux can be restated as a constraint on the average null energy \citep{Myers:2010jv}.
\begin{eqnarray}
\langle {E}_{\hat n} \rangle \sim \langle\int dy^- T_{--}(y^+=0, y^-, y^\alpha)\rangle \geq 0.
\end{eqnarray}
In order to compute the energy functional from holography, one considers fluctuations of the metric $h_{\mu \nu}$ which are dual to the stress tensor insertion at the boundary. We extend coordinates of \eqref{coord} into the bulk of $AdS_D$ and consider the following shock wave background.
\begin{eqnarray}\label{genshock}
ds^2 &=& \frac{\tilde{L}^2}{u^2}\left( \delta(y^+)W(\bar{y},u)(dy^+)^2-dy^+dy^-+d\bar{y}^2 +du^2 \right) ,
\end{eqnarray}
where $u=\frac{z}{x^+}$, $\bar{y}^2 = \sum_{\alpha=1}^{D-3} y_\alpha y^\alpha$ and $z=0$ is the boundary of $AdS_D$. Analogous to the Gauss Bonnet case the scaled curvature is defined as 
\begin{eqnarray}
\tilde{L}^2 &=& \frac{L^2}{f(0)},
\end{eqnarray}
where $f(u)$ is defined in \eqref{genfdef}.
This solves the equations of motion for any higher derivative gravity background provided it satisfies the following equation of motion \cite{Horowitz:1999gf}.
\begin{eqnarray}\label{Weom}
\partial_u^2 W - \frac{D-2}{u}W + \sum_{i=1}^{D-3}  \partial_i^2 W =0.
\end{eqnarray}
The primary motivation behind considering such a background is that the solution for $W(y,u)$ can be used to source the energy functional $E_{\hat{n}}\sim \int dy^- T_{--}(y^+=0, y^-, y^\alpha)$ 
at the boundary \citep{Hofman:2009ug, Myers:2010jv}. The required solution with the appropriate asymptotic behaviour is given by
\begin{eqnarray}\label{wsol}
W(\bar{y},u) &=& \frac{2^{D-2}}{(1+n^{D-2})^{D-2}} \frac{u^{D-1}}{(u^2 + (\bar{y} - \bar{Y})^2)^{D-2}},\nonumber\\
&&\bar{y}=y^\alpha, \qquad \bar{Y}=\frac{n^\alpha}{1+n^{D-2}}.
\end{eqnarray}
where $n^i$ is the $i$ th component of the unit vector $\hat{n}$. We now introduce perturbations corresponding to the operators $\mathcal{O}$ creating the states in \eqref{energyfunc}. 
\begin{eqnarray}\label{genpert}
h_{y^1 y^2} = \frac{\tilde{L}^2}{u^2} \phi(y,u).
\end{eqnarray}
There are other modes accompanying $\phi(y,u)$ in order for the perturbation to be transverse as well as traceless in the bulk, $h^\mu_\mu = \nabla^\mu h_{\mu \nu}=0$. However they do not affect our three point function \cite{Myers:2010jv, Buchel:2009sk}. The component $h_{y^1 y^2}$ satisfies the equation of a massless scalar in $AdS_D$ (ignoring interaction terms with the shock wave).
\begin{eqnarray}\label{phieom}
\partial_u^2 \phi - \frac{D-2}{u}\phi + \sum_{i=1}^{D-3}  \partial_i^2 \phi - 4\partial_{y^+}\partial_{y^-} \phi =0.			
\end{eqnarray}
We plug the shock wave solution \eqref{genshock} with the perturbation \eqref{genpert} into the action  \eqref{actionGEN}. Using the equations of motion of $W$ and $\phi$ (\eqref{phieom} and \eqref{Weom})as well as integration by parts we arrive at, 
\begin{eqnarray}
S_{\phi^2W}&=& -\frac{1}{\kappa}\int d^Dy \sqrt{-g}\delta(y^+)\phi \partial_-^2 \phi W \left(1-4 \kappa \left(c_1 (D-1) D+ c_2 (D-1)-2 c_3(D-4)\right)\right.\nonumber\\
&&\phantom{-\frac{1}{\kappa}\int d^Dy \sqrt{-g}\delta(y^+)\phi \partial_-^2 \phi W }\left. +2c_3 \kappa \left(2 f(0)T_2\right) \right), \nonumber\\
\end{eqnarray} 
where
\begin{eqnarray}
T_2 &=& \frac{\partial_1^2 W +\partial_2^2 W -2\partial_u W}{W}.
\end{eqnarray}
The integral localises on $u=1, y^\alpha =0$ \citep{Buchel:2009sk}. From \eqref{wsol}, we can evaluate $T_2$.
\begin{eqnarray}
T_2 &=& 2(D-1)(D-2)\left(\frac{(n^1)^2+(n^2)^2}{2}-\frac{1}{D-2} \right).
\end{eqnarray}
 Putting everything together and comparing with the known standard form of the energy functional in \eqref{energyfngenform}, we have,
\begin{eqnarray}
S_{\phi^2W} &=& -\frac{\mathcal{C}_T}{\kappa \tilde{\alpha}}\int d^{D-1}y du ~ \sqrt{-g}\delta(y^+)\phi \partial_-^2 \phi W \left( 1 + t_2\left( \frac{(n^1)^2+(n^2)^2}{2}-\frac{1}{D-2} \right)\right),\nonumber\\
\end{eqnarray}  
with
\begin{eqnarray}
t_2 &=& 8c_3\kappa(D-1)(D-2) + O(c_i^2).
\end{eqnarray}
The shear sum rule \eqref{gensumrule} to leading order in coupling constants $c_i$s become,
\begin{eqnarray}
\left(\frac{D-1}{2D}+\frac{-4 c_3 (D-4) (D-1) \kappa}{D-2}\right)\epsilon &=& \lim_{\epsilon \rightarrow 0^+}\frac{1}{\pi}\int_{-\infty}^{\infty} \frac{\delta\rho(z)dz}{z -i\epsilon} .
\end{eqnarray}
We obtain a match with the shear sum rule computed holographically (\eqref{gensumruleholo}). The field theory computation is an independent calculation which relies on the parameters of the three point function of the stress tensor. Therefore, this provides a non trivial check of our sum rule. 

\section{Conclusion}
                 We have derived holographic shear sum rules corresponding to the spectral density of the stress tensor $T_{xy}$ for possible theories which are dual to Einstein gravity with higher derivative corrections, truncating at fourth order in derivatives. We find that to leading order in the coupling constants $c_i$, the holographic shear sum rule gets modified compared to the large $\lambda_c$ limit, which is pure Einstein gravity in $AdS_D$ background. We also compute the shear sum rule from field theory by calculating the dimensionless collider parameters $t_2$ and $t_4$ for the possible dual field theory at finite 't Hooft coupling $\lambda_c$ and obtain a match with the holographic results. We have also obtained the shear sum rule for Gauss Bonnet theories in $D=5$. The computation is exact in the coupling constant $\lambda_{GB}$ and matches with field theory predictions. Our analysis is a stringent check for the coefficient of $t_2$ in the sum rule, derived from conformal invariance. 
                 \paragraph{} While we have proven the analyticity of the retarded correlator for Gauss Bonnet gravity, we have assumed that it remains so for the general case. A formal proof of the analyticity of the retarded correlator for  general curvature squared corrections would involve a non-perturbative black hole solution to the action \eqref{actionGEN}. Moreover, for theories with quadratic curvature corrections, $t_4=0$. Therefore, in the setup we have considered, it is not possible to check the coefficient of $t_4$ in \eqref{introdsumrule}. In order to test the coefficient of $t_4$, we need to incorporate cubic curvature corrections to the classical action as explored in \cite{Oliva:2010eb, Myers:2010ru, Oliva:2011xu}. The authors of \citep{Myers:2010ru} consider a specific combination of cubic curvature corrections to the Einstein Hilbert action such that the equations of motion are second order in derivatives and obtain exact black hole solutions in $AdS$ background. The Hofman Maldacena Coefficient $t_2$ and $t_4$ are non zero for CFTs which are dual to such a quantum corrected action \cite{Myers:2010jv}. It would be interesting to compute shear sum rules for such cubic curvature corrections to two derivative Einstein gravity. This involves obtaining the equations of motion corresponding to the fluctuations in the shear channel and computing the effective action after taking into account the appropriate Gibbons-Hawking terms and the counter terms. We leave these topics as future endeavours.

\acknowledgments
                     The author would like to thank Justin David for countless discussions, encouragement and a very careful reading of the manuscript. The author would also like to thank Christopher Herzog, Juan Maldacena and Kallol Sen for discussions regarding higher derivative gravity theories. The author is grateful to the high energy theory departments in Yale university, YITP Stonybrook, Princeton University, University of Illinois, Urbana Champaign and Enrico Fermi Institute, Chicago for hospitality and the opportunity to present a part of this work.

\appendix
\section{Details of Gauss Bonnet gravity}\label{GBappendix}
                              The details of the gibbons hawking term and the counter terms have been discussed in detail in \cite{Grozdanov:2016fkt}.
\begin{eqnarray}
S_{GH} &=& -\frac{1}{k_5^2} \int d^4x \sqrt{-\gamma} \left[ K + \lambda_{GB}(J- 2G_\gamma^{\mu \nu}K_{\mu\nu}) \right], \nonumber\\
S_{c.t} &=& \frac{1}{k_5^2}\int d^4x \sqrt{- \gamma} (c_1 - \frac{c_2}{2} R_\gamma), \nonumber\\
c_1 &=& -\frac{\sqrt{2}(2+\sqrt{1-4\lambda_{GB}})}{\sqrt{1+\sqrt{1-4\lambda_{GB}}}} \qquad c_2 = \sqrt{\frac{\lambda_{GB}}{2}}\frac{(3-4\lambda_{GB}-3\sqrt{1-4\lambda_{GB})}}{(1-\sqrt{1-4\lambda_{GB})^{\frac{3}{2}}}},\nonumber\\
\end{eqnarray}
where $\gamma_{\mu\nu}=g_{\mu\nu}-n_\mu n_\nu$ is the induced metric on the boundary and $n_\mu$ is a vector normal to the boundary. $R_\gamma$ and $G_\gamma^{\mu \nu}$ are the induced Ricci scalar and Einstein tensor on the boundary respectively. The extrinsic curvature tensor is given by
\begin{eqnarray}
K_{\mu\nu} &=& -\frac{1}{2}(\nabla_\mu n_\nu + \nabla_\nu n_\mu ),
\end{eqnarray} 
$K$ is its trace and the tensor $J_{\mu \nu}$ is given by,
\begin{eqnarray}
J_{\mu \nu} &=& \frac{1}{3}(2KK_{\mu \alpha}K^{\alpha}_\nu + K_{\alpha \beta} K^{\alpha \beta} K_{\mu \nu} -2K_{\mu \alpha} K^{\alpha \beta} K_{\beta \nu}- K^2K_{\mu \nu}).
\end{eqnarray}
 Trace of $J_{\mu \nu}$ is denoted by $J$.
 \paragraph{ }
The terms in the equation of motion of the fluctuation $\delta g_{xy}$ (\ref{GBeom}) is given by,
\begin{eqnarray}\label{compoeomgb}
A(u) &=& -u \left(\frac{1}{\left(\gamma_{GB}^2-1\right) \left(1-u^2\right)^2-u^2+1}+\frac{1}{\left(1-u^2\right) \sqrt{\gamma_{GB}^2-\left(\gamma_{GB}^2-1\right) u^2}}\right)-\frac{1}{u} \nonumber\\
B(u) &=&-\frac{\left(\gamma_{GB}^2-1\right)^2 \lambda ^2 \left(\left(\gamma_{GB}^2-1\right) u^2+U-\gamma_{GB}^2\right)}{8 (\gamma_{GB}+1) u \sqrt{\gamma_{GB}^2-\left(\gamma_{GB}^2-1\right) u^2} \left(U-1\right) \left(\left(\gamma_{GB}^2-1\right) u^2+2 U-\gamma_{GB}^2-1\right)} \nonumber\\ 
\end{eqnarray}
\section{Details of general higher derivative corrections}\label{genappendix} 
    The terms in the equations of motion of the fluctuation $\delta g_{xy}(u)$ is given by
\begin{eqnarray}\label{compoeomgen}
C(u) &=& -\left(\frac{8 \gamma  u \left(1-\frac{1}{2} (D-1) u^2\right)}{(D-3) f(u)}+\frac{2 u}{f(u)}+\frac{1}{u}\right), \nonumber\\
D(u) &=& \frac{4f(0) \omega^2}{r_0^2(D-1)^2f(u)^2 u^{\frac{2(D-3)}{(D-1)}}},
\end{eqnarray}              
where            
\begin{eqnarray}
f(u)=1-u^2+\alpha +\gamma u^4
\end{eqnarray}
The coefficients in the on shell action \eqref{onshellgen} are given by
\begin{eqnarray}\label{compoeffgen}
%\tilde{A}(u) &=& \frac{8 \lambda ^{-\frac{3 D}{2}-\frac{1}{2}} y^{-\frac{D}{2}-\frac{3}{2}} \left(\frac{\alpha +\gamma +D-1}{D-1}\right)^{\frac{1}{2} (-D-1)} }{(D-2) (\alpha +\gamma +D-1)} \left((D-2) (D-1)^2 \lambda ^{D+1} y^{D+1} \left(\frac{\alpha +\gamma +D-1}{D-1}\right)^{D+1} \right. \nonumber\\
%&&\left.(4 k (c_1 (D-1) D+c_2 (D-1)-c_3 (D-5))-1)-y^2 \lambda ^{2 D} (\alpha +\gamma +D-1)^2 \right. \nonumber\\
%&&\left. (2 D k (c_1 (D-1) D+c_2 (D-1)+2 c_3)-D+2)+2 c_3 (D-3) \left(D^2-3 D+2\right)^2\right.\nonumber\\
%&&\left. k \lambda ^2 y^{2 D} \left(\frac{\alpha +\gamma +D-1}{D-1}\right)^{2 D}\right)\nonumber\\
\tilde{A}(u) &=& 16 (D-1) \left((D-1) \kappa \left(2 u-\frac{D}{(D-2) u}\right) (c_1 D+c_2) \right.\nonumber\\
&&\left. +c_3 \kappa \left((D-3) (D-2) u^3-2 (D-5) u-\frac{2 D}{(D-2) u}\right)+\frac{1-u^2}{2 u}\right) \nonumber\\
\tilde{B}(u) &=& (D-1) \left(\frac{6 \left(1-u^2\right)}{u}-\frac{(D-1) \kappa \left(12 c_1 D^2-c_2 \left(D^2-15 D+2\right)\right)}{(D-2) u}\right.\nonumber\\
&&\left. +(D-1) \kappa u (24 c_1 D+2 c_2 (D+11))+c_2 (D-1)^2 \kappa u^3\right. \nonumber\\
&&\left. +c_3 \kappa \left(4 \left(8 D^2-43 D+49\right) u^3-16 \left(D^2-6 D+3\right) u+\frac{4 \left(D^3-8 D^2+19 D-26\right)}{(D-2) u}\right)\right)\nonumber\\
\tilde{E}(u)&=& (D-1)^3 \kappa \left(1-u^2\right)^2 u (c_2+4 c_3)\nonumber\\
\tilde{F}(u)&=& -2 (D-1)^2 \kappa \left(1-u^2\right) \left(c_2 (D-1) \left(u^2+1\right)+4 c_3 \left(2 (D-3) u^2-(D-5)\right)\right) \nonumber\\
\end{eqnarray}
\subsection*{Useful Integrals} 
\begin{eqnarray} \label{integrals}
&&\int_0^\infty dy\phantom{d} y^{D-1} K_{\frac{D-1}{2}}(y) K_{\frac{D-3}{2}}(y) = 2^{D-4} \Gamma \left(\frac{D-1}{2}\right)^2 \nonumber\\
&&\int_0^\infty dy\phantom{d}  y^{D+1} K_{\frac{D-1}{2}}(y) K_{\frac{D-3}{2}}(y) = \frac{\sqrt{\pi } \Gamma (D) \Gamma \left(\frac{D+1}{2}\right)}{4 \Gamma \left(\frac{D}{2}+1\right)} \nonumber\\
&&\int_0^\infty dy\phantom{d}  y^{D} K_{\frac{D+1}{2}}(y) K_{\frac{D-3}{2}}(y) = \frac{2^{D-2} \Gamma \left(\frac{D-1}{2}\right) \Gamma \left(\frac{D+3}{2}\right)}{D} \nonumber\\
&&\int_0^\infty dy\phantom{d}  y^{D+2} K_{\frac{D+1}{2}}(y) K_{\frac{D-3}{2}}(y) = \frac{\sqrt{\pi } \Gamma (D+1) \Gamma \left(\frac{D+5}{2}\right)}{2 (D+1) \Gamma \left(\frac{D}{2}+2\right)} \nonumber\\
&&\int_0^\infty dy\phantom{d}  y^{D} K_{\frac{D-1}{2}}(y)^2 = \frac{\sqrt{\pi } \Gamma (D) \Gamma \left(\frac{D+1}{2}\right)}{4 \Gamma \left(\frac{D}{2}+1\right)} \nonumber\\
&&\int_0^\infty dy\phantom{d}  y^{D+1} K_{\frac{D+1}{2}}(y)K_{\frac{D-1}{2}}(y) = 2^{D-2} \Gamma \left(\frac{D+1}{2}\right)^2
\end{eqnarray} 
\subsection*{Two point function calculation}
                   The coefficients appearing in the on shell action \eqref{gen2ptfnaction} are,
\begin{eqnarray}\label{coeffaction2ptfn}
\tilde{C}(r) &=& -\frac{r^D \left(k \left(4 D (c_1 (D-1) D+c_3 ((D-8) D+27))+c_2 \left(D^3+D-2\right)-104 c_3\right)-2 D+4\right)}{4 \sqrt{f(0)} (D-2) k} \nonumber\\
\tilde{D}(r) &=&-\frac{(D-1) (c_2+4 c_3) r^D}{4 \sqrt{f(0)}}
\end{eqnarray}

\providecommand{\href}[2]{#2}\begingroup\raggedright\endgroup

 \end{document}